\documentclass[a4paper,showpacs,preprintnumbers,amsmath,amssymb,twocolumn]{revtex4}

\usepackage{graphicx}% Include figure files
\usepackage{dcolumn}% Align table columns on decimal point
\usepackage{bm}% bold math
\usepackage{epstopdf}
\newcount\driver
\newcount\bozza

scaled\magstep1                  
scaled\magstep1 
 
scaled\magstep1                 \font\indbf=cmbx10 scaled\magstep2

scaled \magstep2

{\count255=\time\divide\count255 by 60
\xdef\hourmin{\number\count255}
        \multiply\count255 by-60\advance\count255 by\time
   \xdef\hourmin{\hourmin:\ifnum\count255<10 0\fi\the\count255}}

\let\a=\alpha \let\b=\beta    \let\g=\gamma     \let\d=\delta     \let\e=\varepsilon
  \let\h=\eta     \let\th=\vartheta \let\k=\kappa     \let\l=\lambda
\let\m=\mu    \let\n=\nu                      \let\r=\rho
\let\s=\sigma \let\t=\tau            
\let\ps=\psi   \let\o=\omega     
 \let\D=\Delta       \let\L=\Lambda    
             
\let\O=\Omega

\def\RR{{\cal R}}

\def\pp{{\bf p}}\def\xx{{\bf x}}
   
\def\yy{{\bf y}}\def\nn{{\bf n}}
\def\zz{{\bf z}}

       \def\oo{{\underline \omega}}
\def\ee{{\underline \varepsilon}}

\let\==\equiv

\let\io=\infty
\let\0=\noindent

\def\*{{\hfill\break\null\hfill\break}}

\def\tilde#1{{\widetilde #1}}

\def\tende#1{\,\vtop{\ialign{##\crcr\rightarrowfill\crcr
             \noalign{\kern-1pt\nointerlineskip}
             \hskip3.pt${\scriptstyle #1}$\hskip3.pt\crcr}}\,}
\def\otto{\,{\kern-1.truept\leftarrow\kern-5.truept\to\kern-1.truept}\,}

\def\Tr{\rm Tr}

\def\wh#1{\widehat{#1}}
\def\hat#1{\wh{#1}}
\def\sqt[#1]#2{\root #1\of {#2}}

\def\bp{{\bar \ps}}

\def\RR{{\cal R}}

\def\T#1{{#1_{\kern-3pt\lower7pt\hbox{$\widetilde{}$}}\kern3pt}}
\def\VVV#1{{\underline #1}_{\kern-3pt
\lower7pt\hbox{$\widetilde{}$}}\kern3pt\,}
\def\W#1{#1_{\kern-3pt\lower7.5pt\hbox{$\widetilde{}$}}\kern2pt\,}

\def\indica{\leaders \hbox to 0.5cm{\hss.\hss}\hfill}
\def\guida{\leaders\hbox to 1em{\hss.\hss}\hfill}
\mathchardef\oo= "0521

\def\pp{{\bf p}}\def\xx{{\bf x}}
\def\yy{{\bf y}}\def\nn{{\bf n}}
\def\zz{{\bf z}}

\def\oo{{\underline \omega}}

\def\qed{\raise1pt\hbox{\vrule height5pt width5pt depth0pt}}
 
  \def\bp{{\bar p}} 

\def\indic{\hbox{\raise-2pt \hbox{\indbf 1}}}

\def\ins#1#2#3{\vbox to0pt{\kern-#2 \hbox{\kern#1 #3}\vss}\nointerlineskip}

\newdimen\xshift \newdimen\xwidth \newdimen\yshift
\def\insertplot#1#2#3#4#5#6{%
\xwidth=#1pt \xshift=\hsize \advance\xshift by-\xwidth \divide\xshift by 2%
\begin{figure}[ht]
\vspace{#2pt} \hspace{\xshift}
\begin{minipage}{#1pt}
#3 \ifnum\driver=1 \griglia=#6
\ifnum\griglia=1 \openout13=griglia.ps \write13{gsave .2
setlinewidth} \write13{0 10 #1 {dup 0 moveto #2 lineto } for}
\write13{0 10 #2 {dup 0 exch moveto #1 exch lineto } for}
\write13{stroke} \write13{.5 setlinewidth} \write13{0 50 #1 {dup 0
moveto #2 lineto } for} \write13{0 50 #2 {dup 0 exch moveto #1
exch lineto } for} \write13{stroke grestore} \closeout13
\includegraphics{griglia.ps} \fi
\includegraphics{#4.ps}\fi%
\ifnum\driver=2 \fi
\end{minipage}
\caption{#5}
\end{figure}
}
\newdimen\shift \shift=-1.5truecm
\def\lb#1{%
\ifnum\bozza=1
\label{#1}\rlap{\hbox{\hskip\shift$\scriptstyle#1$}}
\else\label{#1} \fi}

\def\be{\begin{equation}}
\def\ee{\end{equation}}
\def\bea{\begin{eqnarray}}\def\eea{\end{eqnarray}}
\def\bean{\begin{eqnarray*}}\def\eean{\end{eqnarray*}}
\def\bfr{\begin{flushright}}\def\efr{\end{flushright}}
\def\bc{\begin{center}}\def\ec{\end{center}}
\def\bal{\begin{align}}\def\eal{\end{align}}
\def\ba#1{\begin{array}{#1}} \def\ea{\end{array}}
\def\bd{\begin{description}}\def\ed{\end{description}}
\def\bv{\begin{verbatim}}\def\ev{\end{verbatim}}
\def\nn{\nonumber}
\def\Halmos{\hfill\vrule height10pt width4pt depth2pt \par\hbox to \hsize{}}
\def\pref#1{(\ref{#1})}

\def\ins#1#2#3{\vbox to0pt{\kern-#2 \hbox{\kern#1 #3}\vss}\nointerlineskip}

\newdimen\xshift \newdimen\xwidth \newdimen\yshift
\newcount\griglia

\def\insertplot#1#2#3#4#5#6{%
\xwidth=#1pt \xshift=\hsize \advance\xshift by-\xwidth \divide\xshift by 2%
\begin{figure}[ht]
\vspace{#2pt} \hspace{\xshift}
\begin{minipage}{#1pt}
#3 \ifnum\driver=1 \griglia=#6
\ifnum\griglia=1 \openout13=griglia.ps \write13{gsave .2
setlinewidth} \write13{0 10 #1 {dup 0 moveto #2 lineto } for}
\write13{0 10 #2 {dup 0 exch moveto #1 exch lineto } for}
\write13{stroke} \write13{.5 setlinewidth} \write13{0 50 #1 {dup 0
moveto #2 lineto } for} \write13{0 50 #2 {dup 0 exch moveto #1
exch lineto } for} \write13{stroke grestore} \closeout13
\includegraphics{griglia.ps} \fi
\includegraphics{#4.ps}\fi%
\ifnum\driver=2 \fi
\end{minipage}
\caption{#5}
\end{figure}
}
\newdimen\shift \shift=-1.5truecm
\def\lb#1{%
\label{#1}\rlap{\hbox{\hskip\shift$\scriptstyle#1$}}
\else\label{#1} \fi}

\def\be{\begin{equation}}
\def\ee{\end{equation}}
\def\bea{\begin{eqnarray}}\def\eea{\end{eqnarray}}
\def\bean{\begin{eqnarray*}}\def\eean{\end{eqnarray*}}
\def\bfr{\begin{flushright}}\def\efr{\end{flushright}}
\def\bc{\begin{center}}\def\ec{\end{center}}
\def\bal{\begin{align}}\def\eal{\end{align}}
\def\ba#1{\begin{array}{#1}} \def\ea{\end{array}}
\def\bd{\begin{description}}\def\ed{\end{description}}
\def\bv{\begin{verbatim}}\def\ev{\end{verbatim}}
\def\nn{\nonumber}
\def\Halmos{\hfill\vrule height10pt width4pt depth2pt \par\hbox to \hsize{}}
\def\pref#1{(\ref{#1})}

\usepackage{amsmath}
\usepackage{amsfonts}
\usepackage{amssymb}
\usepackage{epstopdf}
\usepackage{color}

scaled\magstep1

\let\a=\alpha \let\b=\beta  \let\g=\gamma  \let\d=\delta
\let\e=\varepsilon
  \let\h=\eta   \let\th=\theta \let\k=\kappa \let\l=\lambda
\let\m=\mu    \let\n=\nu             \let\r=\rho
\let\s=\sigma \let\t=\tau    
\let\ps=\Psi   \let\o=\omega
 \let\D=\Delta  \let\L=\Lambda 
         
\let\O=\Omega

\def\RR{{\cal R}}

 \def\pp{{\bf p}}
 \def\xx{{\bf x}} \def\yy{{\bf y}} \def\zz{{\bf z}}

\def\nn{\nonumber}

\def\\{\hfill\break}
\def\={:=}
\let\io=\infty
\let\0=\noindent

\def\tende#1{\,\vtop{\ialign{##\crcr\rightarrowfill\crcr\noalign{\kern-1pt
    \nointerlineskip} \hskip3.pt${\scriptstyle #1}$\hskip3.pt\crcr}}\,}
\def\otto{\,{\kern-1.truept\leftarrow\kern-5.truept\to\kern-1.truept}\,}

\def\wh{\widehat}
\def\to{\rightarrow}

\def\qed{\hfill\raise1pt\hbox{\vrule height5pt width5pt depth0pt}}

\def\be{\begin{equation}}
\def\ee{\end{equation}}
\def\bp{\begin{pmatrix}}
\def\ep{\end{pmatrix}}
\def\bea{\begin{eqnarray}}
\def\eea{\end{eqnarray}}
\def\nn{\nonumber}
\def\pref#1{(\ref{#1})}

\def\lb{\label}

\def\Tr{\mathrm{Tr}}

\begin{document}

\title{Spin Hall insulators beyond the Helical Luttinger model}

\author{Vieri Mastropietro}%
\affiliation{University of Milan, Via Saldini 50, 20133 Milan, Italy}
\author{Marcello Porta}
\affiliation{University of Z\"urich, Winterthurerstrasse 190, 8057 Z\"urich, Switzerland\footnote{New address: Eberhard Karls Universit\"at T\"ubingen, Auf der Morgenstelle 10, 72076 T\"ubingen, Germany}}
\begin{abstract}
We consider the interacting, spin conserving, extended Kane-Mele-Hubbard model, and we rigorously
establish the exact quantization of the edge spin conductance and the validity of the Helical Luttinger liquid relations for Drude weights and susceptibilities. Our analysis fully takes into account lattice effects, typically neglected in the Helical Luttinger model approximation, which play an essential role for universality. The analysis is based on exact renormalization group methods and on a combination of lattice and emergent Ward identities, which allow to relate the emergent chiral anomaly with the finite renormalizations due to lattice corrections.
\end{abstract}

\maketitle

%%%%%%%%%%%%%%%%%%%%%%%%%%%%%%%%%%%%%%%%%%%%%%%%%%%%%%%%

\section{Introduction}

The remarkable edge transport properties of Quantum Spin Hall insulators (QSHI), predicted in \cite{1,2,3,4,5} (see  \cite{3a,3aa,3b} for reviews), have been explained so far via topological arguments or effective Quantum Field Theory (QFT) descriptions. In the absence of many-body interactions, and if the spin is conserved, topological arguments ensure the quantization of the spin Hall conductance. Many-body interactions, however,
break the single-particle picture, and prevent the use of such methods. Nevertheless, experiments have shown values of spin conductances that are approximately quantized \cite{6,7,8,9,10}. It is a challenge for theorists to understand a mechanism for universality, or to the predict possible deviations from the quantized value.

Due to the reduced dimensionality and to the massless dispersion relation, the edge states form a strongly correlated system. In order to analytically understand its behavior, the {\it Helical Luttinger} (HL) {\it model} \cite{4}, a QFT for relativistic one-dimensional fermions with locked spin and chirality, has been proposed as an effective field-theoretic description. This model can be studied via bosonization, see {\it e.g.} \cite{10a,10b}; as a result, it exhibits anomalous decay of correlations, and the {\it chiral anomaly}. Also, nonuniversal anomalous exponents, velocities and transport coefficients are related by {\it exact scaling relations}. Several generalizations of the HL model have been considered, see \cite{11,12,13,14,15,16,17,18,19,20,21,22}. However, these effective QFT descriptions are insufficient to conclude whether many-body interactions break or not the quantization of the spin conductance, since they neglect important lattice effects; it is well-known that nonlinear corrections to the dispersion relation and Umklapp terms might produce finite corrections to the transport coefficients, as for instance in graphene \cite{12h, 13h}.

In this paper we establish, for the first time, the exact quantization of the edge spin conductance of a truly interacting lattice QSHI, going beyond the effective QFT description. Moreover, we establish the validity of the HL scaling relations, by fully taking into account lattice effects and the nonlinearity of the energy bands. We use recently developed nonperturbative RG methods, introduced to prove rigorous universality results for nonsolvable statistical mechanics models, \cite{15aaa}. 

\section{The KMH model}

\noindent{{\it The model.}} A basic model for interacting, time-reversal invariant topological insulators is provided by the 
extended, spin-conserving {\it Kane-Mele-Hubbard (KMH)} model. The KMH model is a time-reversal symmetric system, describing spinful fermions on the honeycomb lattice. The honeycomb lattice $\L$ can be represented as the superposition of two triangular sublattices $\L_{A}, \L_{B}$ of side $L$, $\L = \L_{A} + \L_{B}$. We denote by $\vec \ell_{1}$, $\vec \ell_{2}$ the normalized basis vectors of $\L_{A}$, and we set $\L_{B} = \L_{A} + (1, 0)$. We shall denote by $x_{1}, x_{2}$ the coordinates of the point $\vec x \in \L_{A}$ in the $\vec\ell_{1}, \vec \ell_{2}$ basis. We introduce fermionic creation/annihilation operators $a^{\pm}_{\vec x,\s}$, $b^{\pm}_{\vec y, \s}$, with spin labels $\s = \pm$, acting on the two triangular sublattices $\L_{A}$, $\L_{B}$.
\begin{figure}[hbtp]
\centering
\includegraphics[width=.35
\textwidth]{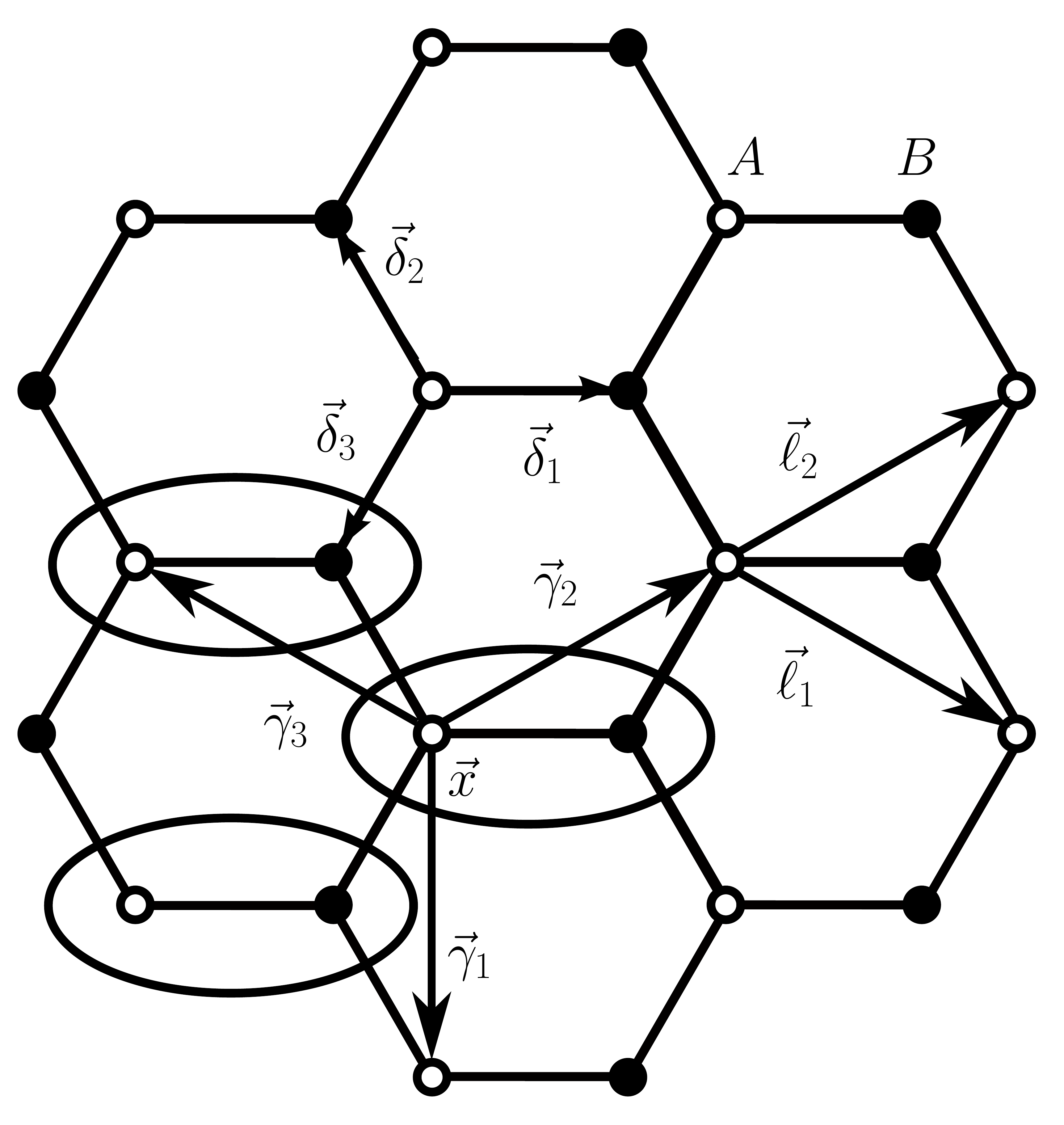}
\caption{The honeycomb lattice $\L$: the empty dots belong to the $A$-sublattice, while the black dots belong to the $B$-sublattice. The ovals encircle the two sites of the fundamental cell, labeled by the position of the empty dot, i.e., of the site of the $A$ sublattice.}\label{fig:haldane}
\end{figure}
In the absence of interactions, the Hamiltonian is:
\bea\label{eq:Ham}
&&\mathcal{H}_{0} = -t_{1}\sum_{\vec x, j, \s} [ a^{+}_{\vec x, \s} b^{-}_{\vec x + \vec \d_{j}, \s} + b^{+}_{\vec x + \vec \d_{j}, \s} a^{-}_{\vec x, \s} ]\nn\\
&& - i t_{2} \Big[ \sum_{\substack{\langle \langle \vec x, \vec y \rangle \rangle\\ \s}} a^{+}_{\vec x,\s} ( \vec \s \vec \n_{\vec x, \vec y} ) a^{-}_{\vec y,\s} + \sum_{\substack{\langle \langle \vec x, \vec y \rangle\rangle \\ \s}}b^{+}_{\vec x,\s} ( \vec \s \vec \n_{\vec x, \vec y} ) b^{-}_{\vec y,\s}\Big]\nn\\
&& - W\sum_{\vec x, \s} [ a^{+}_{\vec x, \s} a^{-}_{\vec x, \s} - b^{+}_{\vec x+\d_{1}, \s} b^{-}_{\vec x + \d_{1}} ] - \m \mathcal{N}
%&&\mathcal{V} = \frac{1}{2}\sum_{\vec x, \vec y} v(\vec x - \vec y) \Big[\rho_{\vec x} - \frac{1}{2}\Big]\Big[ \r_{\vec y}- \frac{1}{2}\Big]\;,\nn
\eea
where in the first sum $\vec x\in \L_{A}$ and $j = 1,2,3$ labels one of its three nearest-neighbours in $\L_{B}$,  connected by the vectors $\vec \d_{j}$, see Fig. \ref{fig:haldane}. The second and third sums run over next-to-nearest neighbours on the $A, B$ sublattices, connected by the vectors $\pm \vec \g_{j}$, $j=1,2,3$; we denote by $\vec \s = (\s_{1}, \s_{2}, \s_{3})$ the vector of the Pauli matrices, and we set
\be 
\vec \n_{\vec x,\vec y}=(\vec d_{\vec x,\vec z}\times \vec d_{\vec z,\vec y})/|\vec d_{\vec x,\vec z}\times \vec d_{\vec z,\vec y}|\;,
\ee 
where $\vec z$ is the intermediate site between $\vec x$ and $\vec y$, and $\vec d_{\vec x,\vec y} = \vec x - \vec y$. The third term includes a staggered potential $\pm W$ on the $A, B$ sublattices, and the last term fixes the chemical potential $\m$ ($\mathcal{N}$ is the number operator). The Hamiltonian of the model is the sum of two copies of the {\it Haldane model} \cite{Ha}: $\mathcal{H}_{0} = \sum_{\s} \mathcal{H}_{0}^{\s}$, where $\mathcal{H}_{0}^{\s}$ acts on the $\s$-spin subsector. The connection between the different spin sectors is $\mathcal{H}^+_0 = C \mathcal{H}^- C$ with $C$ the complex conjugation operator, which ensures the invariance under time-reversal symmetry of the full Hamiltonian.

In order to reduce the honeycomb lattice to a Bravais lattice, we collect the fermionic operators associated to the sites $\vec x$, $\vec x + \vec \d_{1}$ in a single, two-component fermionic operator (see Fig. \ref{fig:haldane}): $\phi^{+}_{\vec x,\s} = (a^{+}_{\vec x, \s}, b^{+}_{\vec x + \vec\d_{1},\s})\equiv (\phi^{+}_{\vec x, A, \s}, \phi^{+}_{\vec x, B, \s})$. With these notations, we rewrite the noninteracting Hamiltonian as:
\be
\mathcal{H}_{0} = \sum_{\vec x,\vec y} \sum_{\r, \r', \s} \phi^{+}_{\vec x, \r, \s} H^{\s}_{\r\r'}(\vec x, \vec y) \phi^{-}_{\vec y, \r', \s}\;,
\ee
with $H^{\s}$ a one-particle Schr\"odinger operator, acting on $\L_{A} \times \mathbb{C}^{2}$. Let us now define the density operator as $\r_{\vec x,\s} = \sum_{\r = A, B} \r_{\vec x, \r, \s}$, with $\r_{\vec x, \r, \s} = \phi^{+}_{\vec x, \r, \s} \phi^{-}_{\vec x, \r, \s}$. The {\it interacting} Hamiltonian is:
\bea\label{H}
\mathcal{H} &=& \mathcal{H}_{0} + \l \mathcal{V}\nn\\
\mathcal{V} &=& \sum_{\vec x, \vec y} \sum_{\r,\r'} \Big[\r_{\vec x,\r,\s} - \frac{1}{2}\Big]\Big[ \r_{\vec y, \r', \s'} - \frac{1}{2}\Big] v_{\r\r'}(\vec x, \vec y)
\eea
for $v_{\r\r'}(\vec x, \vec y)$ short ranged, and where $\l$ is the coupling constant.

\medskip

\noindent{{\it Lattice currents and conservation laws.}} Let $A(t) = e^{i\mathcal{H}t } A e^{-i\mathcal{H}t}$ be the time-evolution of $A$. The density operator satisfies the following lattice continuity equation:
\bea\label{eq:cont0}
\partial_{t} \rho_{\vec x,\s}(t) &=& i[ \mathcal{H}, \rho_{\vec x,\s}(t)] = \sum_{\vec y}\sum_{\r,\r'} j^{\r\r';\s}_{\vec x,\vec y}(t)\nn\\
j_{\vec x, \vec y}^{\r\r';\s} &=& i \phi^{+}_{\vec y, \r} H^{\s}_{\r\r'}(\vec y, \vec x) \phi^{-}_{\vec x, \r'} + \text{h.c.}
\eea
The operator $j_{\vec x, \vec y}^{\r\r';\s}$ is the {\it bond current operator}, corresponding to the pairs of honeycomb lattice sites labelled by $(\vec x,\r;  \vec y, \r')$. Notice that, by the finite range of the hopping Hamiltonian, the only nonvanishing bond currents are those connecting $(\vec x, \vec x \pm \vec \ell_{i})$, with $i = 1,2$, and $(\vec x, \vec x \pm \vec \g_{1})$, with $\vec \g_{1} = \vec \ell_{1} - \vec \ell_{2}$. 

Let $j_{\vec x, \vec y}^{\s} = \sum_{\r,\r'} j^{\r\r';\s}_{\vec x, \vec y}$. Let us define the discrete lattice derivative as: $\text{d}_{i} f(\vec x) = f(\vec x) - f(\vec x - \vec \ell_{i})$. Then, the continuity equation can be rewritten as $\partial_{t} \rho_{\vec x, \s}(t) = $
\bea &&-\text{d}_{1} j_{\vec x, \vec x + \vec \ell_{1}} - \text{d}_{2} j_{\vec x, \vec x + \vec \ell_{2}} - j_{\vec x, \vec x + \vec \ell_{1} - \vec \ell_{2}} - j_{-\vec \ell_{1} + \vec \ell_{2} + \vec x, \vec x}\nn\\&&\equiv
-\text{d}_{1} j_{1, \vec x} - \text{d}_{2} j_{2, \vec x}\;,
\eea
where we defined $j^{\s}_{1,\vec x} = j^{\s}_{\vec x, \vec x+ \vec \ell_{1}} + j^{\s}_{\vec x, \vec x + \vec \ell_{1} - \vec \ell_{2}}$ and $j^{\s}_{2,\vec x} = j^{\s}_{\vec x, \vec x + \vec \ell_{2}} + j^{\s}_{\vec x, \vec x - \vec \ell_{1} + \vec \ell_{2}}$. We shall collect densities and currents in a single $3$-current $j^{\s}_{\m, \vec x}$, $\m = 0,1,2$. Also, we define the charge and spin $3$-currents as $j^{\text{c}}_{\m, \vec x} = \sum_{\s} j^{\s}_{\m, \vec x}$, $j^{\text{s}}_{\m, \vec x} = \sum_{\s} \s j^{\s}_{\m, \vec x}$, which satisfy $\partial_{0} j^{\sharp}_{0, \vec x} + \sum_{i} \text{d}_{i} j^{\sharp}_{i, \vec x}= 0$, with $\sharp = \text{c}, \text{s}$.

We shall study the thermodynamic properties of the model in the grand canonical ensemble. The Gibbs state of the model is $\langle \cdot \rangle_{\beta,L} = (1/ \mathcal{Z}_{\b, L}) \Tr \cdot e^{-\beta \mathcal{H}}$, with $\mathcal{Z}_{\b, L} = \Tr\, e^{-\beta \mathcal{H}}$ the partition function. We introduce
the imaginary-time (or Euclidean) evolution of the fermionic operators as:
$
\phi^{\pm}_{\xx,\r, \s} := e^{x_{0} \mathcal{H}} \phi^{\pm}_{\vec x, \r, \s} e^{-x_{0} \mathcal{H}}$, $\xx = (x_{0}, \vec x)$ with $x_{0}\in [0, \beta)$, extended antiperiodically for all $x_{0}\in \mathbb{R}$. %The zero temperature charge and spin conductivity are defined according to Kubo formula. For convenience, we shall define these transport coefficients after Wick rotating to imaginary times; this step can be justified rigorously, and we refer the reader to \cite{GMP1, M66} for details. The zero temperature charge and spin conductivity matrices are $\s^{\text{c}}_{ij} = \s^{+}_{ij} + \s^{-}_{ij}$, and $\s^{\text{s}}_{ij} = \s^{+}_{ij} - \s_{ij}^{-}$, with $\s^{\pm}_{ij} = $
%
%\be
 %\lim_{p_0\to 0}\lim_{\vec p\to 0}\lim_{\b, L\to\io}
%{1\over p_0}\int_{0}^{\b}dx_{0}\sum_{x\in\L}(1 - e^{-i\pp\cdot \xx})\langle j^\pm_{i,\xx}\,; j^\pm_{j,\bf 0}\rangle_{\b,L}\;.
%\ee
%
A crucial ingredient in our analysis will be the use of {\it Ward identities}, implied by the charge and spin conservation laws. Let $d_{0} \equiv i\partial_{x_{0}}$. The lattice contintuity equation can be rewritten in a compact form as $\sum_{\m}\text{d}_{\m} j^{\s}_{\m,\xx} = 0$. This relation can be used to derive identities among correlations, such as:
\be\label{eq:WI}
\sum_{\m}\text{d}_{x_\m} \langle{\bf T} j^{\s}_{\m,\xx}\,; j^{\s}_{\n,\yy} \rangle_{\beta,L} = i\d(x_{0} - y_{0}) \langle [ j^{\s}_{0, \vec x}\,, j^{\s}_{\n, \vec y} ] \rangle_{\b, L}
\ee
In Eq. (\ref{eq:WI}), {\bf T} is the time-ordering operator, and the contact term in the right-hand side is called the {\it Schwinger term}. Eq. (\ref{eq:WI}) is the Ward identity for the current-current correlation functions. 
In the same way, one can also derive a Ward identity relating the vertex functions of the lattice model to the two point correlation function:
\bea\label{eq:WIvertex}
&&\sum_{\m} \text{d}_{z_\m} \langle {\bf T} j^{\sharp}_{\m, \zz}\,; \phi^{-}_{\yy, \s,\rho'} \phi^{+}_{\xx,\s, \rho} \rangle_{\beta,L} = \\
&&i\s_\sharp \big[\langle {\bf T} \phi^{-}_{\yy ,\s,\rho'} \phi^{+}_{\xx, \s,\rho}\rangle_{\beta,L}\delta_{\xx, \zz} - \langle {\bf T} \phi^{-}_{\yy , \s,\rho'} \phi^{+}_{\xx, \rho}\rangle_{\beta,L}\delta_{\yy,\zz}\big]\nn
\eea
where $\d_{\xx,\zz} = \d(x_{0} - y_{0})\d_{\vec x, \vec y}$ and $\s_c=+$, $\s_s=\s$.

\section{Non-interacting topological insulators}

In the absence of interactions, $\l = 0$, the Hamiltonian reduces to the sum of two noninteracting Haldane Hamiltonians, $\mathcal{H}_0 = \sum_{\s = \pm} \mathcal{H}_{0}^{\s}$. Suppose the model is equipped with periodic boundary conditions. Then, (using that the single-particle Hamiltonian is translation invariant, $H^{\s}(z,z') \equiv H^{\s}(z - z')$) we can introduce the {\it Bloch Hamiltonian} as $\hat H^{\s}(\vec k) = \sum_{z} e^{-i\vec z\cdot \vec k}H^{\s}(z)$, for $\vec k$ in the Brillouin zone $\mathcal{B}$. We have:
\be
\hat H^{\s}(\vec k) = \begin{pmatrix} m_\s(k)    & -t_{1} \Omega^*( k) \\ - t_{1}\Omega( k) & - m_\s(k)   \end{pmatrix}\ee
where
$
m_\s(\vec k) =
 W-2 \s t_{2} \a(\vec k)$, $\a(\vec k) = \sum_{i=1}^3 \sin \vec k\cdot \vec \g_i $, $
\O(\vec k) = 1+e^{-i \vec k\cdot \vec \ell_1}+e^{-i \vec k\cdot \vec \ell_2}$.
The corresponding energy bands are 
\begin{equation}
E_{\pm}^\s(\vec k) = \pm \sqrt{m_\s(\vec k)^{2} + t^{2}|\Omega(\vec k)|^{2}}\;.\nn
\end{equation}

To make sure that the energy bands do not overlap, we assume that $t_2/t_1<1/3$.
The two bands can only touch  at the {\it Fermi points} $\vec k_{F}^{\pm} = \big( \frac{2\pi}{3}, \pm \frac{2\pi}{3\sqrt{3}} \big)$, which are the two zeros of $\Omega(\vec k)$,
around which $\O(\vec k_F^\pm+\vec k')\simeq \frac32(i k_1'\pm k_2')$.
The condition that the two bands touch
at $\vec k_F^\o$, with $\o = +,-$, is that $m^{\s}_\o=0$, with 
\be
m_{\o}^\s \equiv m_\s(\vec k_{F}^{\o}) = W +\o \s 3\sqrt{3}\,t_{2} \;.\nn
\ee
Therefore, the unperturbed critical points are given by the values of $W$ such that $W=\pm 3\sqrt3\, t_2$. Choosing the chemical potential $\m = 0$, which lies halfway between the two energy bands, the condition $W \neq \pm 3\sqrt3\, t_2$ corresponds to the insulating phase, for which the correlations decay exponentially fast. In the insulating phase, the system may or may not be in a topologically nontrivial phase, depending on the value of the {\it Hall conductivity}. This quantity is defined starting from Kubo formula, which we use directly in its imaginary time version (see \cite{GMP1} for a discussion of the Wick rotation): $\s_{12}^\s = $
\be
\lim_{p_0\to 0}\lim_{\vec p\to 0}\lim_{\b, L\to\io}
{1\over p_0}\int_{0}^{\b}dx_{0}\sum_{x}(1 - e^{-i\pp\cdot \xx})\langle j^\s_{1,\xx}\,; j^\s_{2,\bf 0}\rangle_{\b,L}\;.
\ee

In the absence of interactions, the Hall conductivity of the Haldane model can be computed explicitly. One finds:
\be\label{s}
\sigma_{12}^\s = \frac{\nu^{\s}}{2\pi}\;,\qquad \nu^\s = \text{sign}(m^\s_{-}) - \text{sign}(m^\s_{+})\;.
\ee
Concerning the Kane-Mele model, its net Hall conductivity $\s^{\text{c}}_{12} = \s^{+}_{12} + \s^{-}_{12}$ vanishes, while the net {\it spin} conductivity $\s^{\text{s}}_{12} = \s^{+}_{12} - \s^{-}_{12}$ is nonzero:
\be
\s^{\text{s}}_{12}=\s^+_{12}-\s^-_{12}=\frac{\nu^+}{\pi}\;.\label{sigma}
\ee
This is the {\it quantum spin Hall effect}. In the spin-symmetric case, the quantization of $\s^{\text{s}}_{12}$ follows from the quantization of $\s_{12}^{\s}$, which is ensured by topological reasons. In the absence of spin symmetry, for instance in the presence of Rashba couplings, one does not expect the spin conductivity to be quantized. Nevertheless, topology survives in the sense that the Hamiltonians are classified by a suitable $\mathbb{Z}_{2}$ invariant \cite{1,2}. 

A remarkable feature of topological insulators is the presence of {\it gapless edge modes}. Suppose now the system is equipped with {\it cylindric boundary conditions}, say periodic in the $\vec \ell_{1}$ direction and Dirichlet in the $\vec \ell_{2}$ direction, on the boundaries at $x_{2} = 0$, $x_{2} = L$. By translation invariance in the $\vec \ell_{1}$ direction, we can introduce a partial Bloch transformation of the initial Hamiltonian, $\hat H_{\r\r'}(k_{1}; x_{2}, y_{2}) = \sum_{z_{1}} e^{-i z_{1} k_{1}} H_{\r\r'}(z_{1}; x_{2}, y_{2})$, with $k_{1} \in S^{1}$. By construction, the Hamiltonian is symmetric under the action of the time-reversal operator, $T^{*}\hat H(k_{1}) T \equiv T^*( \hat H^{+}(k_{1}) + \hat H^{-}(k_{1}) )T =\overline{\hat H^{-}(-k_{1})} + \overline{\hat H^{+}(-k_{1})} \equiv \hat H(-k_{1})$ since $\overline{\hat H^{\s}(k_{1})} = \hat H^{-\s}(-k_{1})$. {\it Edge states} correspond to solutions of the Schr\"odinger equation $\hat H(k_{1})\xi(k_{1}) = \e(k_{1}) \xi(k_{1})$ at the Fermi level $\m$, which are exponentially localized around one of the two edges: 
\be 
|\xi_{x_{2}}(k_{1})|\leq Ce^{-c|x_{2}|} \quad |\xi_{x_{2}}(k_{1})|\leq Ce^{-c|L-x_{2}|}\;.
\ee
These $1d$ eigenfunctions of $\hat H(k_{1})$ correspond to $2d$ eigenfunctions for the Hamiltonian $H$, of the form $e^{-ik_{1}x_{1}} \xi_{x_{2}}(k_{1})$; they are responsible for the transport of dissipationless {\it edge currents}. 
In the Haldane model the edge eigenfunctions can be found explicitly \cite{Hao}: each cylinder edge supports either zero or one edge mode. Consequently, the Kane-Mele Hamiltonian $H = \sum_{\s = \pm} H^{\s}$ supports either zero or two edge states per edge. Let $\e_{+}$, $\e_{-}$ be their dispersion relations. By time-reversal symmetry, $\e_{+}(k_{1}) = \e_{-}(-k_{1})$: the model displays two Fermi points $k_{F}^{\pm}$, $k_{F}^{+} = -k_{F}^{-}$, such that $$\e_{+}(k_{F}^{+}) = \e_{-}(k_{F}^{-}) = \m\;.$$ Time-reversal symmetry implies that the edge modes are counterpropagating: $v_{+} = \partial_{k_{1}} \e_{+}(k_{F}^{+}) = -v_{-}$.

The edge transport of the system can be investigated by probing the variation of the density or of the current (of charge or of spin) after introducing an external perturbation supported in a strip of width $a$ from the $x_{2} = 0$ edge. We shall study these transport phenomena in the linear response regime. In order to define the edge transport coefficients, let us introduce the following notations. Given a local operator $O_{\vec x}$, we define its partial space-time Fourier transform as $\hat O_{\underline{p}, x_{2}} = \int_{0}^{\beta} dx_{0} \sum_{x_{1}} e^{-i\underline{p}\cdot \underline{x}} O_{\xx}$, with $\underline{p} = (p_{0}, p_{1})$, $p_{0}$ the Matsubara frequency, and $\underline{x} = (x_{0}, x_{1})$. Let $\pmb{\langle} \cdot \pmb{\rangle}_{\infty} = \lim_{\beta, L\to \infty} (\beta L)^{-1} \langle \cdot \rangle_{\b, L} $. We define, for $\sharp, \sharp' = \text{c}, \text{s}$:
\bea\label{eq:Gs}
G_{\r,\r}^{\underline{\sharp};a}(\underline{p}) &=& \sum_{x_{2} = 0}^{a} \sum_{y_{2} = 0}^{\infty} \pmb{\langle} {\bf T} \rho^{\sharp}_{\underline{p}, x_{2}}\,; \rho^{\sharp'}_{\underline{p}, y_{2}} \pmb{\rangle}_{\infty}\nn\\
G_{\r, j}^{\underline{\sharp};a}(\underline{p}) &=& \sum_{x_{2} = 0}^{a} \sum_{y_{2} = 0}^{\infty} \pmb{\langle} {\bf T} \rho^{\sharp}_{\underline{p}, x_{2}}\,; j^{\sharp'}_{1,-\underline{p}, y_{2}} \pmb{\rangle}_{\infty}\label{nat}  \\
G_{j, j}^{\underline{\sharp};a}(\underline{p}) &=& \sum_{x_{2} = 0}^{a} \big[ \sum_{y_{2} = 0}^{\infty} \pmb{\langle} {\bf T} j^{\sharp}_{1,\underline{p}, x_{2}}\,; j^{\sharp'}_{1,-\underline{p}, y_{2}} \pmb{\rangle}_{\infty} - i\D(x_{2})\big]\nn
\eea
where $\D(x_{2}) = \lim_{\beta, L\to \infty} \langle \sum_{\s} [t^{\s}_{\vec x,\vec x + \vec \ell_{1}} + t^{\s}_{\vec x, \vec x + \vec \ell_{1} - \vec \ell_{2}}] \rangle_{\beta, L}$, with 
$t_{\vec x, \vec y}^{\s} = \sum_{\r\r'}-i \phi^{+}_{\vec y, \r} H^{\s}_{\r\r'}(\vec y, \vec x) \phi^{-}_{\vec x, \r'} - \text{h.c.}$. As we shall see later, this function is related to the Schwinger term in Eq. (\ref{eq:WI}). The {\it edge spin conductance} is:
\be
\s^{\text{s}} = \lim_{a\to \infty} \lim_{p_{0}\to 0^+}\lim_{p_{1}\to 0} G^{\text{c,s};a}_{\r, j}(\underline{p})\;.
\ee
It measures the variation of the spin current after introducing a shift of the chemical potential supported in a region of width $a$ from the $x_{2} = 0$ edge. Similarly, the {\it edge charge conductance} is:
\be
\s^{\text{c}} =  \lim_{a\to \infty} \lim_{p_{0}\to 0^+}\lim_{p_{1}\to 0} G^{\text{c,c};a}_{\r, j}(\underline{p})\;.
\ee 
Instead, the edge susceptibilities and Drude weights, of charge or of spin, are:
\bea 
\k^\sharp &=& \lim_{a\to\io}\lim_{p_1\to 0}\lim_{p_0\to 0^+}  G^{\sharp, \sharp; a}_{\r, \r}(\underline{p})\nn\\
D^\sharp &=& -\lim_{a\to\io}\lim_{p_0\to 0^+}\lim_{p_1\to 0}  G^{\sharp, \sharp; a}_{j, j}(\underline{p})\;.
\eea
As we shall see, due to the lack of continuity at $\underline{p} = (0,0)$ of the expressions in (\ref{eq:Gs}), the order of the limits in the above definitions is crucial. It turns out that, in the absence of interactions, the edge transport coefficients can be computed. One has:
\bea
&&\s^{\text{c}} = 0\;,\qquad \s^{\text{s}} = \s_{12}^{\text{s}}\;,\nn\\
&& \kappa^{\sharp} = \frac{1}{\pi |v_{+}|}\;,\qquad D^{\sharp} = \frac{|v_{+}|}{\pi}\;.\label{lll}
\eea
The equivalence of the edge spin conductance with the bulk spin conductivity
 is a manifestation of the {\it bulk-edge correspondence}: namely, a duality between the presence of edge modes at the Fermi level with the value of the topological invariant classifying bulk Hamiltonians (acting on infinite lattices, with no edges). For the IQHE \cite{Hal, Hat, SKR, EG}, this duality implies that the sum of the chiralities of the edge states
$\sum_e \o_e$, with 
$\o_{e} = \text{sign}( \partial_{k_{1}} \e_{e}(k_{F}^{e}) )$ equals the Chern number of the Bloch bundle, which fixes the value of the Hall conductivity.
 For time-reversal invariant systems, instead, $\frac{1}{2}\sum_{e} |\o_{e}|$ mod $2$ turns out to be equal to the bulk $\mathbb{Z}_{2}$ invariant \cite{QWZ, AVS, GP}; in particular, for the spin-conserving Kane-Mele model, this implies that the edge spin conductance equals the bulk spin conductivity. The bulk-edge correspondence has been rigorously established for single-particle Hamiltonians: there is no general argument ensuring its validity for interacting many-body systems. Finally, notice that in contrast to $\s^{\text{s}}$, the edge susceptibility $\kappa^{\sharp}$ and the Drude weight $D^{\sharp}$ are nonuniversal quantities, depending on the velocity of the edge modes.

The goal of this paper is to understand the effect of many-body interactions of the edge transport coefficients: the natural question we address here is whether some form of universality persists, and in particular if the quantization of $\s^{\text{s}}$ holds true.

\section{Main result}

Here we shall consider the edge transport properties of the Kane-Mele-Hubbard model, $\l\neq 0$. Our main result is the following theorem.

\medskip
{\bf Theorem.} {\it Consider the KMH Hamiltonian \pref{H} with cylindric boundary conditions. Let us choose the chemical potential $\m$ in the gap of the bulk Hamiltonian. Suppose that the single-particle KM Hamiltonian supports a pair of edge modes, $\e_{+}(k_{F}^{+}) = \e_{-}(k_{F}^{-}) = \m$, and that $v_{+}\neq 0$. Then, there exists $\l_{0}>0$ such that, for $|\l| < \l_{0}$, the following is true. Let $\o = \text{sgn}(v_+)$. The edge spin conductance is universal:
\be
\s^{s} = -{\o \over \pi}\;.\label{ma}
\ee
Moreover, the Drude weights and the susceptibilities satisfy the Helical Luttinger liquid relations:
\be\label{eq:kappaD}
\k^{c} ={K \over \pi v}\;,\quad D^{c}= {v K \over \pi}\;, \quad \k^s= {1 \over \pi v K}\;, \quad D^s={ v \over \pi K}
\ee
with $K = 1 + O(\l) \neq 1$, $v = v_{+} + O(\l) \neq v_{+}$. Finally, the $2$-point function decays with anomalous exponent $\eta = (K+K^{-1}-2)/2$.
}
\medskip

As a corollary, our result combined with the universality of bulk transport, following from the analysis of \cite{GMP1}, \cite{GMP2}, provides the first rigorous example of bulk-edge correspondence for an interacting time-reversal invariant topological insulator (see \cite{M66} for the analogous result for Hall systems).
The lack of many-body corrections to the conductance is in agreement with experimental results, \cite{6, 8}. Notice that, in contrast with the conductance, the susceptibilities and the Drude weights are interaction dependent: nevertheless, if combined with the dressed Fermi velocity $v$, they verify a marginal form of universality, in the sense of the validity of the Helical Luttinger liquid relation:
\be\label{hel}
\frac{\kappa^{\sharp} v^{2}}{D^{\sharp}} = 1\;.
\ee
Moreover, the HL parameter $K$ allows to determine the anomalous exponent of the two-point function, via the formula $\eta = (K+K^{-1}-2)/2$.

The rest of the paper is organized as follows. In Section \ref{sec:1d} we introduce a Grassmann integral representation for the transport coefficients. We then integrate out the ``bulk degrees of freedom'', corresponding to the energy modes far from the Fermi level. As a result, we end up with an effective one-dimensional model, which reminds of the Helical Luttinger model up to some crucial differences: the fermionic fields are defined on a lattice, the interaction involves arbitrarily high monomials in the fields, the energy dispersion relation is nonlinear and Umklapp scattering process are present. Then, in Section \ref{sec:RG} we study this lattice QFT via exact RG, which allows to represent the transport coefficients in terms of renormalized, convergent series. Such expansions can be reorganized by isolating the contributions corresponding to an emergent, effective chiral QFT theory with suitably fine tuned bare parameters, from a remainder term, that depends on all lattice details. The advantage of this rewriting is that the current-current correlation functions of the emergent QFT can be exactly computed, see Section \ref{sec:CLL}, thanks to the validity of extra chiral Ward identities. This allows to compute the edge transport coefficients of the KMH model, up to finite multiplicative and additive renormalizations, dependent on all microscopic details of the model. The values of these renormalizations are, however, severely constrained from one side by the validity of the Adler-Bardeer anomaly nonrenormalization property of the emergent chiral theory, and from the other side by the lattice WIs of the KMH model. As we show in Section \ref{sec:univ}, these facts imply nonperturbative relations among all finite renormalizations, from which our theorem follows.

\section{Reduction to an effective $1d$ theory}\label{sec:1d}

For simplicity, we shall directly consider the case $L=\infty$, which corresponds to having just one edge. 
%The case $L<\infty$ can be studied with minor modifications, as in \cite{M66}.
 It is useful to switch to a functional integral representation of the correlation functions of the lattice model. We define the generating functional of the correlations as:
\be
e^{\mathcal{W}(A)}=\int P(d\Psi)\, e^{-V(\Psi) + B(\Psi;A)}\label{mar} 
\ee
where: $\Psi^\pm_{\xx,\s,\r}$ are Grassman variables, labelled by $\xx = (x_{0}, \vec x)\in [0,\beta)\times \L_{A}$, $\s = \pm$, $\r = A,B$; $P(d\psi)$ is a Gaussian Grassmann integration with propagator given by the noninteracting Euclidean two-point function
\be
g_{\s,\s'}(\xx,\yy) = \d_{\s\s'} \int \frac{d\underline{k}}{(2\pi)^{2}}\, \frac{e^{-i\underline{k}\cdot (\underline{x} - \underline{y})}}{-ik_{0}+  \hat H^{\s}(k_{1}) - \m}(\vec x; \vec y)
\ee
where $\underline{k} = (k_{0}, k_1)$ with $k_{0}$ the fermionic Matsubara frequency and $k_1$ the quasi-momentum associated to the translation invariant direction $\vec \ell_{1}$. The Grassmann counterpart of the many-body interaction is:
\be
V(\Psi) = \l \sum_{\substack{\r, \r' \\ \s, \s'}}\int d\xx d\yy\, n_{\xx,\r,\s} n_{\yy,\r',\s'} v_{\r\r'}(\vec x, \vec y) \delta(x_{0} - y_{0}) \nn
\ee
where $\int d\xx = \int_{0}^{\beta} dx_{0} \sum_{\vec x}$ and $n_{\xx,\r,\s}$ the Grassmann counterpart of the density operator. Finally, $B(\Psi; A)$ is a source term, of the form:
\be
B(\Psi; A) = \sum_{\m, \sharp}\int d\xx\, A_{\m,\xx}^{\sharp} J_{\m,\xx}^{\sharp}
\ee
with $J_{\m,\xx}^{\sharp}$ the Grassmann counterpart of $j^{\sharp}_{\m,\xx}$.

\medskip

We now use the addition principle of Grassmann variables to write $\Psi = \Psi^{\text{(e)}} + \Psi^{\text{(b)}}$, with $\Psi^{\text{(e)}},\, \Psi^{\text{(b)}}$ independent Grassmann variables, with propagators $g^{\text{(edge)}}$, $g^{\text{(bulk)}}$, where $g^{\text{(e)}}$ takes into account the energy modes close enough to the Fermi level. That is: $g^{\text{(e)}}_{\s\s'}(\xx, \yy) =$
\be\label{eq:gedge}
\d_{\s\s'} \sum_{e}\int \frac{d\underline{k}}{(2\pi)^{2}}\, e^{-i\underline{k}\cdot (\underline{x} - \underline{y})}\frac{\chi_{\s}(k_{1})}{-ik_{0} + \e_{\s}(k_1) - \m} P^{\s}_{k_{1}}(x_{2}; y_{2})\;,
\ee
with: $P^{\s}_{k_{1}} = |\xi^{\s} \rangle \langle \xi^{\s}|$, where $\xi^{\s}$ the edge mode of $\hat H^{\s}(k_{1})$, with energy $\e_{\s}$, and $\chi_{\s}(k_{1}) \equiv \chi(|k_{1} - k_{F}^{\s}|\leq \d)$ is a compactly supported cutoff function. By construction, the propagator $g^{\text{(bulk)}}$ is gapped; it only depends on the energy modes that are at a distance at least $\sim \delta$ from the Fermi level. Thus, $|g^{\text{(bulk)}}(\xx,\yy)|\leq Ce^{-c|\xx - \yy|}$. Instead, due to the fact that, for $k_1 = k_1' + k_{F}^{\s}$ and $k_1'$ small
\be
\e_{\s}(k'_{1} + k_{F}^{\s}) - \m = \s v_{+} k'_{1} + O({k'_{1}}^2)\;,
\ee
the edge propagator in Eq. (\ref{eq:gedge}) only decays as $|\underline{x} - \underline{y}|^{-1} e^{-c(|x_{2}| + |y_{2}|)}$. 

The field $\Psi^{\text{(b)}}$ can be integrated out, expanding the integrand of (\ref{mar}) in the coupling $\l$ and using the exponential decay of the bulk propagator together with fermionic cluster expansion techniques \cite{Bry}. We then get:
\be
e^{\mathcal{W}(A)} = e^{\mathcal{W}^{\text{(b)}}(A)} \int P_{\text{e}}(d\Psi^{\text{(e)}}) e^{-V^{\text{(e)}}(\Psi^{\text{(e)}}) + B^{\text{(e)}}(\Psi^{\text{(e)}}; A)}
\ee
where the new effective interaction $V^{\text{(e)}}(\Psi^{\text{(e)}})$ is a sum over monomials $P$ in the fields $\Psi^{(\text{e})}$  of any order $|P| = n$, with kernels $W^{\text{(e)}}_{P}(\xx_{1}, \ldots, \xx_{n})$, exponentially decaying in $|\xx_{i} - \xx_{j}|$ for $i\neq j$. Graphically, a given kernel can be represented as a sum of Feynman diagrams with $|P|$ external lines, corresponding to the edge fields, and an arbitrary number of quartic vertices connected by the bulk propagators. This expansion turns out to be {\it convergent} for small $\l$, thanks to determinant bounds for fermionic field theories, combined with the good decay properties of the bulk propagators. 
%See \cite{M66} for more details in a similar case. 
The new effective source term $B^{\text{(e)}}$ admits a similar representation, where now external lines corresponding to the $A$ fields are present as well.

Due to the special form of the edge propagator, given by Eq. (\ref{eq:gedge}), we now notice that the edge field can be represented as the convolution of a truly one-dimensional field with the edge modes eigenfunctions. That is:
\bea\label{eq:2d1d}
&&\int P_{\text{e}}(d\Psi^{\text{(e)}}) e^{-V^{\text{(e)}}(\Psi^{\text{(e)}}) + B^{\text{(e)}}(\Psi^{\text{(e)}}; A)} \nn\\&& \quad = \int P_{\text{1d}}(d\psi) e^{-V^{\text{(e)}}(\psi*\check\xi) + B^{\text{(e)}}(\psi*\xi; A)}\;,
\eea
where: $P_{\text{1d}}$ is a Grassmann Gaussian integration for a one-dimensional field $\psi^{\pm}_{\vec x,\s}$, with propagator given by, in momentum space:
\be
\hat g_{\s,\s'}(\underline{k}) = \d_{\s\s'}\frac{\chi_{\s}(\underline{k})}{-ik_{0} + \e_{\s}(k_{1}) - \m_{0}}
\ee
where now $\chi_{\s}(|\underline{k}|) = \chi(|\underline{k} - \underline{k}_{F}^{\s}|\leq \d)$, and $\m - \m_{0} = \n_{0}$, with $\n_{0} = O(\l)$ a counterterm, that is chosen so to fix the value of the interacting chemical potential; and 
\be\label{eq:conv}
(\psi^{+} * \check{\xi})_{\xx, \rho}=\sum_{y_{1}} \psi^{-}_{(x_{0}, y_{1}),\s}\, \overline{\check{\xi}^{\s}_{x_{2}}(x_{1} - y_{1}; \rho)}
\ee
where $\check{\xi}^{\s}_{x_{2}}(x_{1};\rho)$ is the Fourier transform of $\chi_{\s}(k_{1})\xi^{\s}_{x_{2}}(k_{1}; \rho)$. This representation of the edge field allows to decouple the $x_{2}$ variables from the remaining $x_{0}, x_{1}$ variables in the effective interaction. Summing over $x_{2}$ (recalling the exponential decay of the edge modes), one finally gets:
\be
e^{\mathcal{W}(A)}=e^{\mathcal{W}^{\text{(b)}}(A)} \int P_{0}(d\psi)\, e^{-V^{(0)}(\psi) + B^{(0)}(\psi;A)}\label{mar1}
\ee
where $P_{0}\equiv P_{\text{1d}}$ and for suitable new effective interaction and source terms, that can be again expressed as sums over monomials of arbitrary order in the $1d$ fields $\psi$. One has: $V^{(0)}(\psi) = $
\be
\int d\underline{x}\, [\l_{0}\psi^{+}_{\underline{x},+}\psi^{-}_{\underline{x},+}\psi^{+}_{\underline{x},-}\psi^{-}_{\underline{x},-} + \sum_{\s}\n_{0} \psi^{+}_{\underline{x},\s} \psi^{-}_{\underline{x},\s}]+\RR V^{(0)}(\psi)
\ee
where the new coupling constant is $\l _{0} = $
\bea
&&\l\sum_{\substack{x_{2}, y_{2} \\ \r,\r'}} \hat v_{\r\r'}(0; x_{2}, y_{2}) \overline{\xi^{(1,\s)}_{x_{2}}(k_{F}; \r)} \xi^{(1,\s)}_{x_{2}}(k_{F}; \r) \nn\\&&\quad\cdot \overline{\xi^{(1,\s)}_{y_{2}}(k_{F}; \r')} \xi^{(1,\s)}_{y_{2}}(k_{F}; \r') + O(\l^2)\;,
\eea
and $\mathcal{R} V^{(0)}$ collects all the higher order terms, together with nonlocal terms. All these contributions turn out to be {\it irrelevant} in the RG sense. Similarly,
\be\label{eq:B0}
B^{(0)}(\psi; A) = \sum_{\m, \sharp}\int d\xx\, Z^{\sharp}_{\m}(x_{2}) A_{\m,\xx}^{\sharp} n^{\sharp}_{\m,\underline{x}}+\RR B^{(0)}(\psi; A)
\ee
where: $Z^{\sharp}_{\m}(x_{2})$ is such that $|Z^{\sharp}_{\m}(x_{2})|\leq Ce^{-cx_{2}}$, and it is analytic in $\l$; and
\bea\label{eq:B00}
&&n^{\text{c}}_{0,\underline{x}} = \sum_{\s} \psi^{+}_{\underline{x},\s}\psi^{-}_{\underline{x},\s}\;,\qquad  n^{\text{c}}_{1,\underline{x}} = \sum_{\s} \s \psi^{+}_{\underline{x},\s}\psi^{-}_{\underline{x},\s}\;,\nn\\
&&\qquad\qquad n^{\text{s}}_{0,\underline{x}} = n^{\text{c}}_{1,\underline{x}}\;,\qquad n^{\text{s}}_{1,\underline{x}} = n^{\text{c}}_{0,\underline{x}}\;.
\eea
Let us give a quick proof of Eqs. (\ref{eq:B0}), (\ref{eq:B00}). After the integration of $\Psi^{\text{(b)}}$ and the reduction to a $1d$ theory, the effective source term has the form, in momentum space: $B^{(0)}(\psi; A) =$
\bea
&&\sum_{\m, \sharp, x_{2}} \int \frac{d\underline{k}}{(2\pi)^{2}}\frac{d\underline{p}}{(2\pi)^{2}}\, \hat A_{\m, (\underline{p}, x_{2})}^{\sharp} \hat \psi^{+}_{\underline{k}+\underline{p}, \s} \hat \psi^{-}_{\underline{k}, \s} \hat W_{\m, \s}^{\sharp}(\underline{p}, \underline{k}; x_{2})\nn\\&&\quad + O(A^{2})\;,
\eea
for suitable kernels $\hat W_{\m, \s}^{\sharp}$. The higher orders in $A$ turn out to be irrelevant in the RG sense. Let us localize the kernel, by writing $\hat W_{\m, \s}^{\sharp}(\underline{p}, \underline{k}; x_{2}) = \hat W_{\m, \s}^{\sharp}(\underline{0}, \underline{k}_{F}^{\s}; x_{2}) + \mathcal{R} \hat W_{\m, \s}^{\sharp}$, where the $\mathcal{R}$ error terms are irrelevant. The effective $1d$ model is invariant under time-reversal symmetry (recall that $\overline{\xi^{\s}(k_{1})} = \xi^{-\s}(-k_{1})$, and that $\hat H^{\s}(k_{1}) = \overline{\hat H^{-\s}(-k_{1})}$):
\be\label{eq:TRS}
\hat A^{\sharp}_{\m, \underline{p}, x_{2}} \to \gamma_{\sharp} \gamma_{\m} \hat A^{\sharp}_{\m, -\underline{p}, x_{2}}\;,\quad \hat \psi^{\e}_{\underline{k}, \s} \to \hat \psi^{\e}_{-\underline{k}, -\s}\;,\quad c\to \bar c\;,
\ee
with $c$ a generic constant in the action, $\gamma_{\text{c}} = 1 = -\gamma_{\text{s}}$ and $\gamma_{0} = 1 = -\gamma_{1}$. This symmetry implies that $\hat W^{\sharp}_{\s,\m}(\underline{0}, \underline{k}_{F}^{\s}; x_{2}) = \gamma_{\sharp}\gamma_{\m} \overline{\hat W^{\sharp}_{-\s, \m}(\underline{0}, \underline{k}_{F}^{-\s}; x_{2})}$. Also, the model is invariant under complex conjugation:
\bea\label{eq:CC}
&&\hat A^{\sharp}_{\m, \underline{p}, x_{2}} \to \hat A^{\sharp}_{\m, \underline{\tilde p}, x_{2}}  \;,\qquad \hat \psi^{+}_{\underline{k}, \s} \to -\hat \psi^{-}_{\underline{\tilde k}, \s}\;,\nn\\&&\quad\qquad \hat \psi^{-}_{\underline{k}, \s}\to \hat \psi^{+}_{\underline{\tilde k}, \s}\;,\qquad c\to \bar c\;.
\eea
with $\tilde k = (-k_{0}, k_1)$. This last symmetry implies that $\hat W^{\sharp}_{\s,\m}(\underline{0}, \underline{k}_{F}^{\s}; x_{2})$ is real. Going back to configuration space, Eq. (\ref{eq:B00}) follows.

Eq. (\ref{mar1}) is an {\it exact} (but very involved) representation of the generating functional the KMH model, in terms of an effective one-dimensional field. It differs from the HL model by the presence of nonlinear corrections in the dispersion and irrelevant terms in the effective interaction. 
%An explicit computation of the correlations via bosonization is thus impossible. Instead, rigorous RG methods %allow to prove the following result. 
%V lo levo perche' supera 4 pg

\section{Multiscale analysis of the edge modes}\label{sec:RG}

Due to the absence of a mass gap, the field $\psi$ cannot be integrated in a single step. Instead, we proceed in a multiscale fashion, exploiting a renormalization procedure at every step. We rewrite the $\psi$ field in terms of single-scale quasi-particle fields, as follows:
\be
\psi^{\pm}_{\underline{x}, \s} = e^{\pm i k^{\s}_{F} x_{1}} \sum_{h = h_{\b}}^{0} \psi_{\underline{x}, \s}^{(h)}
\ee
where each field varies on a scale $2^{-h}$, with $h\leq 0$. The last scale $h_{\beta}$ is fixed by the inverse temperature, $h_{\beta} \sim |\log_{2} \beta|$. The covariance of the fields is defined inductively. We integrate the fields in an iterative fashion. From a RG point of view, the $\psi^{+}_{\underline{x}}\psi^{-}_{\underline{x}}$ terms are relevant, while the $\psi^{+}_{\underline{x}}\partial_{\m} \psi^{-}_{\underline{x}}$, $\psi^{+}_{\underline{x}, \s}\psi^{-}_{\underline{x}, \s} \psi^{+}_{\underline{x}, \s'}\psi^{-}_{\underline{x}, \s'}$ terms are marginal. 

After the integration of the scales $h+1, \ldots, 0$, we obtain the following representation of the generating functional: $e^{\mathcal{W}(A)} = $
\be\label{mar2}
e^{\mathcal{W}^{(h)}(A)} \int P_{h}(d\psi^{(\leq h)}) e^{- V^{(h)}(\sqrt{Z_{h}}\psi) + B^{(h)}(\psi; A)}\;,
\ee
where the new Gaussian Grassmann integration has propagator:
\be
g_{\s,\s'}^{(\le h)}(\underline x,\underline y)= {\d_{\s\s'}\over Z_{h}} \int \frac{d\underline{k}'}{(2\pi)^{2}} {e^{-i\underline k'\cdot  (\underline x-\underline y)}\chi_{h}(\underline k')
\over -ik_0+ \s v_{h} k'_{1}}(1 + r_{h}(\underline{k}'))\label{aa}\nn
\ee
with: $\chi_{h}$ a smooth cutoff function supported for $|\underline{k}'|\leq 2^{h+1}$; $r_{h}$ an error term, $|r_{h}(\underline{k}')|\leq C|\underline{k}'|$; $Z_{h}$, $v_{h}$ the wave function renormalization and the effective Fermi velocity, whose RG flow, as functions of $h$, is marginal. Time-reversal symmetry (\ref{eq:TRS}) and complex conjugation (\ref{eq:CC}) imply that these parameters are real and spin-independent.

The new effective interaction is a sum of Grassmann monomials of arbitrary order. We rewrite it is $V^{(h)} = \mathcal{L} V^{(h)} + \mathcal{R} V^{(h)}$, where $\mathcal{L} V^{(h)}$ takes into account all the relevant and marginal contributions: $\mathcal{L} V^{(h)}(\sqrt{Z_{h}}\psi) =$ 
\be \int d\underline x\, \big[ \l_h Z_{h}^{2} \psi^+_{\underline{x},+}\psi^-_{\underline{x},+}\psi^+_{\underline{x},-}\psi^-_{\underline{x},-} +
\sum_{\s}2^h Z_{h} \n_{h} \psi^+_{\underline{x},\s}\psi^-_{\underline{x},\s}\big]\;, \nn
\ee
while $\mathcal{R} V^{(h)}$ takes into account all irrelevant terms. By the symmetries (\ref{eq:TRS}), (\ref{eq:CC}), the parameters $\l_{h}$, $\n_{h}$ are again real, and spin independent. In the same spirit, we rewrite $B^{(h)} = \mathcal{L} B^{(h)} + \mathcal{R} B^{(h)}$, where $\mathcal{L} B^{(h)}$ collects all marginal terms (there are no relevant terms in the source term):
\be
\mathcal{L} B^{(h)}(\psi; A) = \int d\xx\, Z^{\sharp}_{h,\m}(x_{2}) A_{\m,\xx}^{\sharp} n^{\sharp}_{\m,\underline{x}}\;,
\ee
for suitable (real) running coupling functions $Z_{h, \m}^{\sharp}(x_{2})$.

\medskip

Let us briefly discuss the flow of the running coupling constants. The (relevant) flow of $\n_{h}$ is controlled via a fixed point argument, by properly choosing the initial shift of the chemical potential $\n_{0}$; see \cite{M66} for details in a similar case. Instead, the (marginal) flows of $\l_{h}, v_{h}$, is controlled using a highly nontrivial cancellation in the renormalized expansions, the {\it vanishing of the beta function} \cite{15aaa}, giving $\l_{h} = \l_0 + O(\l^2)$ and $v_{h} = v_{0} + O(\l)$ {\it uniformly} in $h$. Instead, the flows of the wave function and vertex renormalizations {\it diverge} with anomalous exponents, 
\be
Z_{h} \sim 2^{-\eta h}\;,\qquad Z^{\sharp}_{h,\m}(x_{2}) \sim 2^{- \eta h} Z_{0, \m}^{\sharp}(x_{2})\;,
\ee
with $\eta = {\l_0^2\over 8\pi^2 v_0^2} + O(\l_{0}^{4})$.

\medskip

The outcome of this construction is a {\it convergent} expansion for the correlation functions in terms of the running coupling constants, which can be used to prove bounds for the decay of the current-current correlations. Convergence follows from the use fermionic cluster expansion at every step of integration, as in \cite{M66}, and excludes nonperturbative effects. We have:
\be
| \lim_{\b, L\to \infty} \langle {\bf T}\, j^{\sharp}_{\m, \xx}\,; j^{\sharp'}_{\n, \yy} \rangle_{\b, L}|\leq Ce^{-c|x_{2} - y_{2}|}/(1 + |\underline{x} - \underline{y}|^{2})
\ee
This estimate, however, is not  for the computation of the edge transport coefficients. In fact, it is not even enough to prove the boundedness of the Fourier transform of the current-current correlation, uniformly in $\underline{p}$. In order to improve on this, we need to exploit {\it cancellations} in the renormalized expansion, following from the emergent {\it chiral symmetry} of the theory.

\section{Emergent chiral QFT}\label{sec:CLL}

In this section we introduce an emergent effective chiral QFT theory, defined 
by the generating functional: $e^{\mathcal{W}^{\chi}(A)} =$
\be\label{eq:chiralmod}
\int P_N(d\psi) e^{-\l^\chi Z^{\chi2}\int d\underline x d\underline y\, v(\underline x-\underline y) n_{\underline x,+} n_{\underline y,-}+B(\psi; A)}
\ee
where $P_{N}(d\psi)$ is a Gaussian Grassmann measure with propagator:
\be 
g_{\s,\s'}^{\chi}(\underline x, \underline y)={\d_{\s\s'}\over Z^{\chi}}\int \frac{d\underline k}{(2\pi)^2}\,  e^{-i\underline k \cdot (\underline x-\underline y)}  {\chi_N(\underline k)\over -i k_0+\s v^{\chi}  k_{1}}\;;\label{za1o1}
\ee
$\chi_{N}$ is an ultraviolet cutoff, supported for $|\underline{k}|\leq 2^{N+1}$, for $N\gg 1$ (to be sent to infinity at the end). The source term is $B(\psi; A) = \sum_{x_{2}=0}^{\infty}\int d\underline{x}\, Z^{\sharp, \chi}_{\m}(x_2) A^{\sharp}_{\m,\xx} n^{\sharp}_{\m,\underline x}$. The interaction potential $v(\underline{x} - \underline{y})$ is nonlocal and short-ranged. The presence of the UV cut-off is crucial to give a nonperturbative meaning to Eq. (\ref{eq:chiralmod}). Its final removal is done through an ultraviolet multiscale analysis, in which the nonlocal, short-range nature of the interaction plays an essential role \cite{15aaa}. 
The infrared regime of this QFT can be studied as for the lattice model. Let us denote by $\l_{h}^{\chi}$, $Z_{h}^{\chi}$, $v_{h}^{\chi}$, $Z_{\m, h}^{\sharp, \chi}(x_{2})$ the running coupling constants of the emergent chiral model.
$\pmb{\langle}. \pmb{\rangle}^{\chi}$.

The bare parameters $Z^{\chi}, v^{\chi}, \l^{\chi}, Z^{\sharp,\chi}_{\m}$ will be chosen in such a way that the running coupling constants of lattice and chiral theory converge to the same limit as $h\to -\infty$. This fact, together with the convergence of the renormalized expansions for both models, implies that the correlations of the KMH model can be written in terms of the correlations of the emergent chiral model, up to finite multiplicative and additive renormalizations, depending on all microscopic details of the KMH model:
\bea\label{eq:decomp}
&&
\pmb{\langle} {\bf T}\, j^{\sharp}_{\m,\underline{p},x_{2}} j^{\sharp'}_{\n, -\underline{p}, y_{2}} \pmb{\rangle}_{\infty} = \label{eq:latref}\\
&&Z^{\sharp,\chi}_{\m}(x_2) Z^{\sharp,\chi}_{\n}(y_2) \pmb{\langle} n^{\sharp}_{\m,\underline p} n^{\sharp'}_{\n,-\underline p} \pmb{\rangle}^{\chi}
+\hat H^{\sharp, \sharp'}_{\m,\n}(\underline{p}, x_{2}, y_{2})\nn
\eea
where $\pmb{\langle}\cdot \pmb{\rangle}^{\chi}$ denotes the correlations of the emergent chiral model, and $\hat H_{\m,\n}^{\sharp, \sharp'}(\underline{p}; x_{2}, y_{2})$ is an error term, {\it continuous} in $\underline{p}$, in contrast to the first term in the r.h.s. of \pref{eq:latref}. The improved regularity of this contribution is due to the fact that it involves irrelevant terms in the RG sense, which all come with a dimensional gain: in configuration space, such term decays as, for large distances, $e^{-c|x_{2} - y_{2}|}/(1 + |\underline{x} - \underline{y}|^{2+\th} )$ for some $\th>0$. Thus, even though this term disappears pointwise in the scaling limit of the correlations, it gives a {\it finite} contribution to the Fourier transform of the lattice correlations. Concerning the multiplicative renormalizations, they verify the bound $|Z^{\sharp,\chi}_{\m}(x_2)| \leq Ce^{-cx_{2}}$, as a consequence of the exponential decay of the edge states.  

\begin{figure}[hbtp]
\centering
\includegraphics[width=.47
\textwidth]{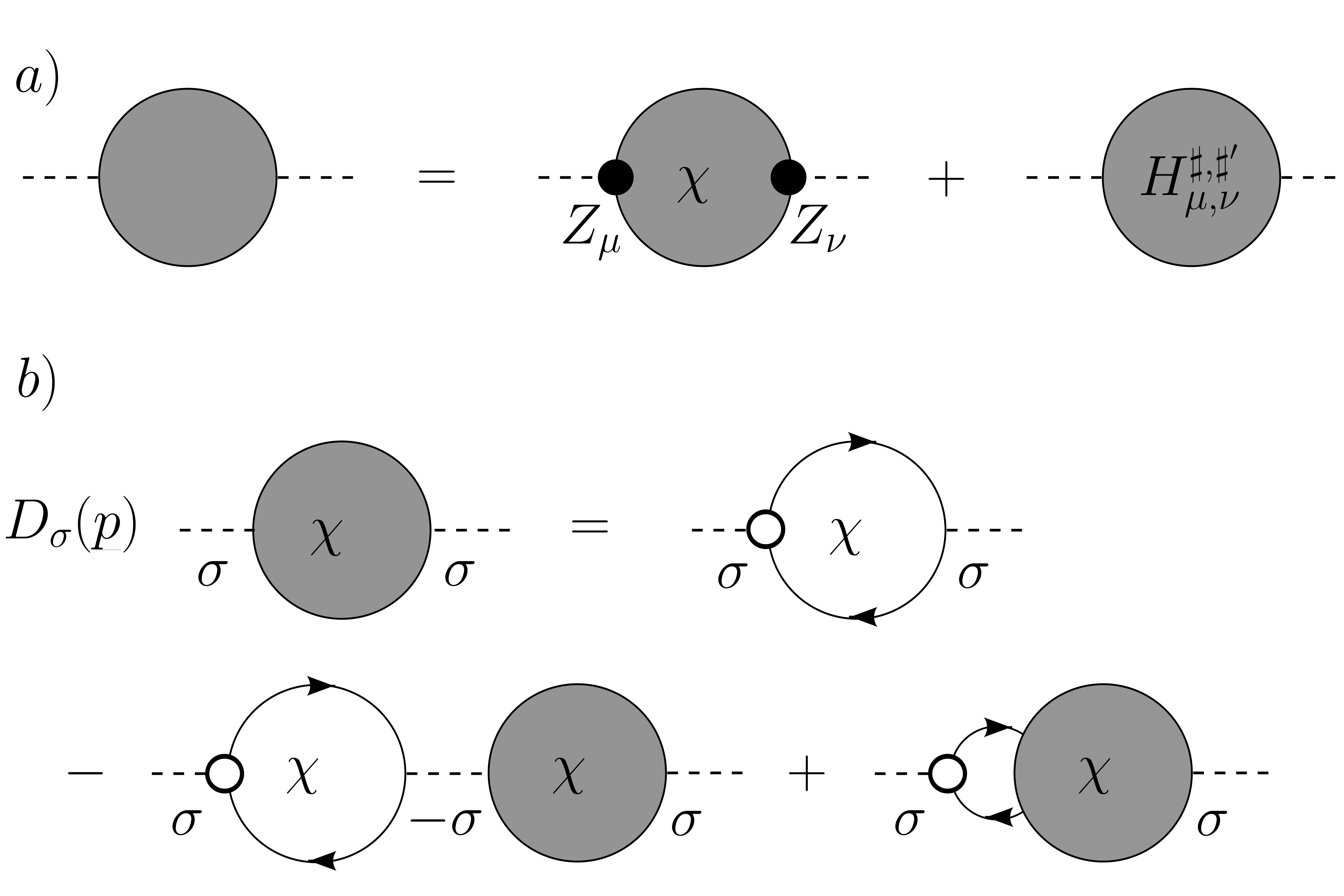}
\caption{$a)$: Graphical representation of Eq. (\ref{eq:latref}). ``$\chi$'' denotes the contributions due to the emergent chiral model. The full dots correspond to the vertex renormalizations, associated to the factors $Z^{\sharp, \chi}_{\m}$ in Eq. (\ref{eq:decomp}). $b)$ Graphical representation of the first WI in Eq. (\ref{eq:Wian}), for a finite UV cutoff $N$. The small white circle denotes a {\it correction vertex}, corresponding to the insertion of $C_{\s}(\underline{p}, \underline{k}) \psi^{+}_{\underline{k}+\underline{p},\s} \psi^{-}_{\underline{k},\s}$. The empty bubble is a noninteracting diagram, whose $N\to \infty$ value is $-\frac{1}{4\pi |v| Z^{2}} D_{-\s}(\underline{p})$.}\label{fig:an}
\end{figure}

Similarly, up to subleading terms for small external momenta:
\bea\label{eq:lat2HL}
&&\pmb{\langle} {\bf T}\, \hat j_{\underline{p},z_{2},\m}^\sharp\,; \hat \phi^-_{\underline{k} + \underline{p}, x_{2}, \r,\s} \hat \phi^+_{\underline{k},y_{2}, \r, \s} \pmb{\rangle}_{\infty}=\label{w111}\\
&&\quad Z^{\sharp,\chi}_{\m}(z_2) Q^{\s}_{x_{2}}(k_{F}^{\o}; \r) \overline{Q^{\s}_{y_{2}}(k_{F}^{\o}; \r)} \pmb{\langle} n^{\sharp}_{\m,\underline p}\,; 
\hat \psi^-_{\underline k+\underline p,\s}
\hat \psi^+_{\underline k,\s} \pmb{\rangle}^{\chi}\nn\eea 
for some functions $Q^{\s}$, such that $Q^{\s} = (1 + O(\l))\xi^{\s}$, which satisfy the exponential bound $|Q^{\s}_{x_{2}}|\leq Ce^{-c|x_{2}|}$. Moreover, up to subleading terms in the external momenta:
\bea
&&\pmb{\langle} {\bf T}\, \hat \phi^-_{\underline{k},x_{2},\r,\s}  \hat \phi^+_{\underline{k}, y_{2}, \r',\s} \pmb{\rangle}_{\infty}\nn\\&&\quad = 
Q^{\s}_{x_{2}}(k_{F}^{\o}; \r) \overline{Q^{\s}_{y_{2}}(k_{F}^{\o}; \r')}\pmb{\langle} \hat \psi^-_{\underline{k},\s}  \hat \psi^+_{\underline{k},\s} \pmb{\rangle}^{\chi}\;.\label{w222}\eea
The advantage of comparing the lattice correlations with those of the emergent model is that the latter can be computed in a closed form, thanks to chiral Ward identities, following from $U(1)$ chiral gauge symmetry. Notice that this symmetry is only approximate, due to the presence of the ultraviolet cutoff. As a result, the UV regularization produces extra terms in the Ward identities of the emergent chiral theory, which do not vanish as $N\to \infty$, but rather produce {\it anomalies} breaking the conservation of the chiral current: see Fig. \ref{fig:an}b). In the figure, the white circle corresponds to the insertion of a {\it correction vertex} $C_{\s}(\underline{p}, \underline{k}) = (\chi_{N}^{-1}(\underline{k}) - 1) D_{\s}(\underline{k}) - (\chi_{N}^{-1}(\underline{k} + \underline{p}) - 1) D_{\s}(\underline{k} + \underline{p})$, with $D_\s(\underline p)= -ip_0+ \s v^{\chi} p_{1}$. For $\underline{p} = O(1)$, this vertex insertion fixes the momentum of the incoming and outgoing fermionic lines on the scale of the ultraviolet cutoff. In the figure, we isolated the terms where the fermionic lines incident to the correction vertex meet at the same point; instead, the last term in the right-hand side of Fig. \ref{fig:an}b) collects all contributions corresponding to diagrams where the lines meet at different points. It turns out that this last term {\it vanishes} as $N\to \infty$, thanks to the nonlocality of the interaction, and to the support properties of the correction vertex. See \cite{Mas, 15aaa} for a detailed proof of this statement, in a similar case.

Setting $D_\s(\underline p)= -ip_0+ \s v^{\chi} p_{1}$, we have:
\bea\label{eq:Wian}
D_{\s}(\underline{p}) \pmb{\langle} \hat \r_{\underline{p}, \s}\,; \hat \r_{-\underline{p},\s} \pmb{\rangle}^{\chi} &=& -\frac{D_{-\s}(\underline{p})}{4\pi |v^{\chi}|{Z^{\chi}}^2}  \\&&+ \tau D_{-\s}(\underline{p}) \hat v(\underline{p}) \pmb{\langle} \hat \r_{\underline{p}, -\s}\,; \hat \r_{-\underline{p}, \s} \pmb{\rangle}^{\chi}\nn\\
\pmb{\langle} \hat \r_{\underline{p}, \s}\,; \hat \r_{-\underline{p}, -\s} \pmb{\rangle}^{\chi} &=&  \tau \frac{D_{-\s}(\underline{p})}{D_{\s}(\underline{p})} \hat v(\underline{p}) \pmb{\langle} \hat \r_{\underline{p}, -\s}\,; \hat \r_{-\underline{p}, -\s} \pmb{\rangle}^{\chi}\nn
\eea
where $\tau = \frac{\l^{\chi}}{4\pi |v^{\chi}|}$ is the {\it chiral anomaly}. The linearity of the anomaly in the bare coupling constant is a highly nontrivial fact, known as {\it Adler-Bardeen anomaly nonrenormalization}. The explicit value of the anomaly can be used to determine the critical exponents of the emergent chiral model. For instance, the anomalous exponent of the two-point Schwinger function is $\h=K+K^{-1}-2$ with $K = \frac{1 - \tau}{1 + \tau}$.

Thus, supposing that $\hat v(\underline{0}) = 1$, we have, up to subleading terms in $\underline{p}$:
\bea\label{sas}
\pmb{\langle} \hat \r_{\underline p,\s} \hat \r_{-\underline p,\s} \pmb{\rangle}^{\chi} &=& {-1\over 4 \pi |v^{\chi}| Z^{\chi2}}{1\over 1-\t^2}
{D_{-\s}(\underline p)\over D_\s(\underline p)}\nn\\
\pmb{\langle} \hat \r_{\underline p,-\s} \hat \r_{-\underline p,\s} \pmb{\rangle}^{\chi} &=& \frac{-1}{4\pi |v^{\chi}| Z^{\chi2} } \frac{\t}{1 - \t^2}\;.
\eea
These expressions can be plugged in the representation for the lattice current-current correlation function, (\ref{eq:decomp}). All we are left to do is to determine the unknown multiplicative and additive renormalizations.

%
%\insertplot{520}{79}
%{\ins{70pt}{40pt}{$hel$}
%\ins{30pt}{40pt}{$D_s(p)$}
%\ins{95pt}{40pt}{$-$}
%\ins{100pt}{40pt}{${\l_\io\over 4\pi} D_{-s}(p)$}
%\ins{155pt}{40pt}{$hel$}
%\ins{190pt}{40pt}{$={D_{-s}(p)\over 4\pi}$}
%}
%{figjsp46}
%{\label{h2} Ward Identity with the anomaly of the chiral model
%} {0}
%

\section{Universality}\label{sec:univ}

In order to fix the values of the finite multiplicative and additive renormalizations we use again Ward identities, this time for the lattice model. These identities introduce nonperturbative relations between the renormalization coefficients, which, as we shall see, imply a dramatic cancellation in the final expression of the edge transport coefficients. To begin, it is convenient to rewrite the Schwinger term of the lattice WI (\ref{eq:WI}) in the following more explicit way:
\bea\label{eq:Sc}
&&\langle [ j^{\s}_{0, \vec x}\,, j^{\s}_{1, \vec y} ] \rangle = (\d_{\vec x, \vec y} - \d_{\vec x, \vec y + \vec \ell_{1}}) \langle t^{\s}_{\vec y,\vec y + \vec \ell_{1}} + t^{\s}_{\vec y, \vec y + \vec \ell_{1} - \vec \ell_{2}} \rangle\nn\\&&\qquad + (\d_{\vec x, \vec y + \vec \ell_{1}} - \delta_{\vec x, \vec y + \vec \ell_{1} - \vec\ell_{2}})\langle t^{\s}_{\vec y, \vec y + \vec \ell_{1} - \vec \ell_{2}} \rangle\nn\\
&&\langle [ j^{\s}_{0, \vec x}\,, j^{\s}_{2, \vec y} ] \rangle = (\d_{\vec x, \vec y} - \d_{\vec x, \vec y + \vec \ell_{2}}) \langle t^{\s}_{\vec y,\vec y + \vec \ell_{2}} + t^{\s}_{\vec y, \vec y - \vec \ell_{1} + \vec \ell_{2}} \rangle\nn\\&&\qquad + (\d_{\vec x, \vec y + \vec \ell_{2}} - \delta_{\vec x, \vec y - \vec \ell_{1} + \vec\ell_{2}})\langle t^{\s}_{\vec y, \vec y - \vec \ell_{1} + \vec \ell_{2}} \rangle
\eea
with $t_{\vec x, \vec y}^{\s} $ defined after \pref{nat}.
Summing up  (\ref{eq:WI}) over $y_{2}$ one gets:
\bea\label{eq:latWI}
&&\text{d}_{y_{0}}\sum_{y_{2}} \langle {\bf T}\, j^{\sharp}_{1, \vec x}\,;  j^{\sharp}_{0, \vec y} \rangle + \text{d}_{y_{1}} \sum_{y_{2}}\langle {\bf T}\, j^{\sharp}_{1, \vec x}\,;  j^{\sharp}_{1, \vec y}\rangle\nn\\&& \quad = i\d(x_{0} - y_{0}) (\d_{x_{1}, y_{1}} - \d_{x_{1}, y_{1} + 1})\D(x_{2})\\
&&\text{d}_{y_{0}}\sum_{y_{2}} \langle {\bf T}\, j^{\sharp}_{0, \vec x}\,;  j^{\sharp'}_{0, \vec y} \rangle + \text{d}_{y_{1}} \sum_{y_{2}}\langle {\bf T}\, j^{\sharp}_{0, \vec x}\,;  j^{\sharp'}_{1, \vec y}\rangle = 0\;.\nn
\eea
To get these relations, we crucially used that $\sum_{y_{2}} \text{d}_{y_{2}} (\cdots) = 0$, which is implied by the Dirichlet boundary conditions. By going into Fourier space, we can use the relations (\ref{eq:latWI}) to prove identities for the edge transport coefficients:
\bea\label{eq:WIlat}
-i p_{0} G^{\sharp, \sharp; a}_{j, \r}(\underline{p}) + p_{1} \eta(p_{1}) G^{\sharp, \sharp; a}_{j,j}(\underline{p}) &=& 0\nn\\
-i p_{0} G^{\sharp, \sharp'; a}_{\r, \r}(\underline{p}) + p_{1} \eta(p_{1}) G^{\sharp, \sharp'; a}_{\r,j}(\underline{p}) &=& 0\;,
\eea
with $p_{1}\eta(p_{1}) = p_{1} + O(p^2_{1})$ the Fourier symbol associated to the lattice derivative $\text{d}_{y_{1}}$. Eqs. (\ref{eq:WIlat}) can be used to determine the $\underline{p}\to \underline{0}$ limit of the additive renormalization $\sum_{x_{2} = 0}^{a} \sum_{y_{2} = 0}^{\infty} \hat H_{\m,\n}^{\sharp,\sharp'}(\underline{p}; x_{2}, y_{2})$ (which exists by continuity in $\underline{p}$). For instance, consider the edge charge conductance, $G^{\text{c}, \text{s}; a}_{\r, j}(\underline{p})$. We can rewrite the second of Eq. (\ref{eq:WIlat}) as $G^{\text{c}, \text{s}; a}_{\r, j}(\underline{p}) = (ip_{0}/ p_{1}\eta(p_{1})) G^{\text{c}, \text{s}; a}_{\r, \r}(\underline{p})$; thus, this relation implies that $\lim_{p_{1}\to 0}\lim_{p_{0}\to 0} G^{\text{c}, \text{s}; a}_{\r, j}(\underline{p}) = 0$. This identity together with the representation (\ref{eq:decomp}) of the current-current correlation function allows to compute the $\underline{p}\to 0$ limit of $\sum_{x_{2} = 0}^{a} \sum_{y_{2} = 0}^{\infty} \hat H_{0,1}^{\text{c}, \text{s}}(\underline{p}; x_{2}, y_{2})$ in terms of the other unknown renormalized parameters. A similar strategy can be followed for the other transport coefficients. 

For simplicity, let us drop the $\chi$ label, and let us set $Z^{\sharp}_{\m} \equiv \sum_{z_2} Z^{\sharp,\chi}_{\m}(z_2)$. The above mentioned strategy allows to compute, up to subleading terms in $\underline{p}$:
\bea\label{eq:G2}
\lim_{a\to \infty}G^{\text{c}, \text{s}; a}_{\r,j}(\underline p) &=& - \frac{Z_{0}^{c} Z^{s}_{1}}{Z^{2} (1 - \t^2)} \frac{1}{\pi |v|} \frac{p_{0}^2}{p_{0}^{2} + v^{2} p_{1}^2}\nn\\
\lim_{a\to \infty}G^{\sharp, \sharp; a}_{j,j}(\underline p) &=& - \frac{Z^{\sharp}_{1} Z^{\sharp}_{1}}{ Z^2 (1 - \t^2) } \frac{1}{\pi |v|} \frac{p_0^2}{p_0^2 + v^2 p_{1}^2}\nn\\
\lim_{a\to \infty}G^{\sharp, \sharp; a}_{\r,\r}(\underline p) &=& {Z^{\sharp}_{0} Z^{\sharp}_{0} \over Z^2 (1-\t^2)}
{1\over \pi |v|}{v^2 p_1^2\over p_0^2 + v^2 p_1^2}\;.
\eea
It remains to determine the multiplicative renormalization in Eqs. (\ref{eq:G2}). This is done by comparing the vertex WIs of lattice and emergent models. From Eq. (\ref{eq:WIvertex}) we have, setting $\eta_{0}(p_1) = -i$:
\bea\label{www}
&&\sum_{\m=0}^1 \eta_{\m}(p_1) \sum_{z_2} \langle {\bf T}\, \hat j_{\underline{p},z_{2},\m}^\sharp\,; \hat \phi^-_{\underline{k} + \underline{p}, x_{2}, \s} \hat \phi^+_{\underline{k},y_{2}, \s} \rangle_{\b,L} = \\
&& \s_{\sharp}[\langle {\bf T}\, \hat \phi^-_{\underline{k},x_{2},\s}  \hat \phi^+_{\underline{k}, y_{2},\s}\rangle_{\b,L} - \langle {\bf T}\, \hat \phi^-_{\underline{k}+\underline{p}, x_{2},\s}  \phi^+_{\underline{k}+\underline{p},y_{2},\s} \rangle_{\b, L}]\nn
\eea
with $\s_{c} = 1$ and $\s_{s} = \s$. On the other hand, the WIs for the emergent chiral model are:
\bea
&& -ip_0 \pmb{\langle} \hat n^{\sharp}_{0,\underline{p}}\,;  \hat \psi^-_{\underline k+\underline p,\s}
\hat \psi^+_{\underline k,\s} \pmb{\rangle} + p_1 v \pmb{\langle} n^{\sharp}_{1,\underline{p}}\,;  \hat \psi^-_{\underline k+\underline p,\s}
\hat \psi^+_{\underline k,\s} \pmb{\rangle} \nn\\
&& = {\s_{\sharp}\over Z (1-\eta_{\sharp} \t)} \big[ \pmb{\langle} \hat \psi^-_{\underline k,\s}
\hat \psi^+_{\underline k,\s}\pmb{\rangle} - \pmb{\langle} \hat \psi^-_{\underline k + \underline{p},\s}
\hat \psi^+_{\underline k + \underline{p},\s} \pmb{\rangle} \big]\label{www1}
\eea
with $\eta_{c} = +,\, \eta_{s} = -$. As before, we now express the lattice correlation functions appearing in the lattice WI in terms of those of the emerging chiral model, using Eqs. (\ref{eq:lat2HL}), (\ref{w222}); we therefore get {\it two} identities for the correlations of the emergent chiral model, one involving the $Z^{\sharp}_{\m}$ parameters, the other involving $Z, v, \tau$. Therefore, we can use these identities to prove relations among these coefficients; we get:
\be
\frac{v Z^{\sharp}_{0}}{Z_{1}^{\sharp}} = 1\;,\qquad \frac{Z^{\sharp}_{0}}{Z(1 - \eta_{\sharp} \tau)} = 1\;.\label{eq:rel}
\ee
Remarkably, Eq. (\ref{eq:rel}) provides a link between the emergent chiral anomaly and the finite lattice renormalizations. We can now use Eq. (\ref{eq:rel}) to simplify the expressions in Eqs. (\ref{eq:G2}). Setting $K^{c}= K$, $K^{s} = K^{-1}$, we get:
\bea
&&\frac{Z^{\sharp}_{0} Z^{\sharp}_{1}}{Z^{2}(1 - \t^2) v} = K^{\sharp}\;,\quad \frac{Z_{0}^{c} Z_{1}^{s}}{Z^2 (1 - \t^2) v} = 1\;,\nn\\
&&\frac{Z^{\sharp}_{1} Z^{\sharp}_{1}}{Z^{2}(1 - \t^2) v} = K^{\sharp}v\;,\quad \frac{Z_{0}^{\sharp} Z_{0}^{\sharp}}{Z^{2}(1 - \t^2) v} = \frac{K^{\sharp}}{v}\;.
\eea
The second relation implies the quantization of $\s^{s}$ (for $\l$ small, $\text{sgn}(v)$ is independent of $\l$). The last two imply the nonuniversality of $D^{\sharp}$, $\kappa^{\sharp}$, and the Helical Luttinger liquid relation $D^{\sharp} = v^{2} \kappa^{\sharp}$. %This concludes the sketch of the proof of the theorem.
%
%\bea
%&&
%{{\bar Z}_{0}^C {\bar Z}_{1}^S\over Z^2(1-\t^2)v}=1 \quad {{\bar Z}_{1}^C {\bar Z}_{1}
%^C\over Z^2(1-\t^2)}=v^2 K
%\nn\\
%&&{{\bar Z}_{1}^S {\bar Z}_{1}^S\over Z^2(1-\t^2)}=v^2/K \quad {{\bar Z}_{1}^C {\bar Z}_{1}^C\over Z^2(1-\t^2)}=
%K\nn\eea
%The first of the above relations says that
%the perfect quantization of the spin edge conductance is consequence of a relation between the chiral anomaly
%of the emerging theory and the renormalization of the currents due to irrelevant terms 
%(charge and spin renormalization are different).  
%The other relations imply the validity
%of helical Luttinger liquid relations for the KMH model; in particular $D_i=v_i^2\k_i$, and the transport coefficients are related to the critical exponents. 
%
\section{Conclusions}

We have established, for the first time, the exact quantization of the edge spin conductance for the spin conserving Kane-Mele-Hubbard model. As a corollary, our result gives an example of bulk-edge correspondence for nonsolvable, interacting time-reversal invariant system. In addition, we proved a marginal form of universality for the susceptibilities and the Drude weights, showing the validity of the Helical Luttinger liquid scaling relations for the KMH model. Our strategy is based on an exact RG construction of the lattice model, and on the combination of lattice Ward identities, following from lattice conservation laws, with relativistic Ward identities, following from the emergent chiral gauge symmetry of the system. Even though they break the integrability of the interacting system, lattice effects and bulk degrees of freedom play a crucial role for universality. 

As an open problem, it would be interesting to include spin nonconserving terms in the Hamiltonian, and to quantify the possible breaking of universality of the edge spin conductance. 
\medskip
\vskip.3cm
{\it Acknowledgements.} V. M. has received funding from the European Research Council (ERC)
under the European Union's Horizon 2020 research and innovation programme (ERC CoG UniCoSM, grant agreement n.724939) and from the Gruppo Nazionale di Fisica Matematica (GNFM). 
The work of M. P. has been partially supported by the NCCR SwissMAP, and by the SNF grant ``Mathematical Aspects of Many-Body Quantum Systems.'' 

%(in the time reversal case) but our method is general. The result
%holds
%for interactions in the convergence domain
%of the series  (whose radius is proven to be finite); in addition, we show the validity of the Luttinger liquid relations
%netween edge exponents, whose validity even at moderate coupings has been questioned \cite{3b}. The proof is based on three key ingradients: a)an exact RG analyis
%of the model, keeping all the irrelevant terms coming from the lattice or backscattering, which contribute to the esge conductivity; b)the validity of lattice Ward Identities posing severe %constraints to the contribution of the irrelevant terms to the edge tansport; 
%z)the emergence of a chiral invariance with associated emerging chiral Ward Identity with a chiral anomaly which verify the Adler-Bardeen non renormalization property. 
%Note that the universality of the conductance is a non pertutative property which could not be establshed by an order by order analysis; the non perturbative information follows from the %Adler-Bardeen non renormalization property, saying that the anomaly is exactly known (is not expressed by a series) and from the constraints posed by the lattice Ward Identities.
%Our result rules out the possibility of corrections in the conductance or in the Luttinger liquid relations provided that the interaction in in the domain of convergence, and it can be used as %a 
%bechmark for nmerical or approximate approaches.


\begin{thebibliography}{9}

\bibitem{1}
 C. L. Kane and E. J. Mele, Phys. Rev. Lett. 95, 226801 (2005)

\bibitem{2}
C. L. Kane and E. J. Mele, Phys. Rev. Lett. 95, 146802 (2005)

\bibitem{3}
B. A. Bernevig and S.-C. Zhang, Phys. Rev. Lett. 96, 106802 (2006)

\bibitem{4}
C. Wu, B. A. Bernevig, and S.-C. Zhang, Phys. Rev. Lett. 96, 106401 (2006)

\bibitem{5} 
C. Xu and J. E. Moore, Phys. Rev. B 73, 045322 (2006)

\bibitem{3a} M. Z. Hasan and C. L. Kane, Rev. Mod. Phys. 82, 3045 (2010)

\bibitem{3aa}
X.-L. Qi and S.-C. Zhang, Rev. Mod. Phys. 83, 1057 (2011)

\bibitem{3b} M. Hohenadler and F. F. Assaad, J. Phys.: Condens. Matter 25, 143201 (2013)

\bibitem{6}
M. K\"onig, S. Wiedmann, C. Br\"une, A. Roth, H. Buhmann,
L. W. Molenkamp, X.-L. Qi, S.-C. Zhang, Science 318, 766
(2007)

\bibitem{7}
A. Roth, C. Br\"une, H. Buhmann, L. W. Molenkamp, J. Maciejko, X.-L. Qi, and S.-C. Zhang, Science 325, 294
(2009)

\bibitem{8}
 K. C. Nowack, E. M. Spanton, M. Baenninger, M. K\"onig,
J. R. Kirtley, B. Kalisky, C. Ames, P. Leubner, C. Br\"une,
H. Buhmann, L. W. Molenkamp, D. Goldhaber-Gordon,
and K. A. Moler, Nat. Mater. 12, 787 (2013)

\bibitem{9}
I. Knez, R.-R. Du, and G. Sullivan, Phys. Rev. Lett. 107,
136603 (2011).

\bibitem{10}
T. Li, P. Wang, H. Fu, L. Du, K. Schreiber, X. Mu, X. Liu, G. Sullivan, G. A. Cs\'athy, X. Lin, R. R. Du, Phys. Rev. Lett. 115 136804 (2015)

\bibitem{10a} 
D. C. Mattis. {\it The Many-Body Problem: An Encyclopedia
of Exactly Solved Models in One Dimension.} World Scientific, Singapore, 1993.

\bibitem{10b} D.C. Mattis, V. Mastropietro. {\it The Luttinger model: the first fifty years and some new directions.} World Scientific, Singapore, 2015.

\bibitem{11}
T. L. Schmidt, S. Rachel, F. von Oppen, and L. I. Glazman, Phys. Rev. Lett. 108, 156402 (2012)

\bibitem{12}
N. Lezmy, Y. Oreg, and M. Berkooz, Phys. Rev. B 85, 235304 (2012)

\bibitem{13} N. Kainaris, I. V. Gornyi, S. T. Carr, and A. D. Mirlin, Phys. Rev. B 90, 075118 (2014)

\bibitem{14}
Y.-Z. Chou, A. Levchenko, and M. S. Foster, Phys. Rev. Lett. 115, 186404 (2015)

\bibitem{15}
A. Strom, H. Johannesson, and G. I. Japaridze, Phys. Rev. Lett. 104, 256804 (2010)

\bibitem{17}
J. Maciejko, C. Liu, Y. Oreg, X.-L. Qi, C. Wu, and S.-C. Zhang, Phys. Rev. Lett. 102, 256803 (2009)

\bibitem{18}
M. Hohenadler and F. F. Assaad, Phys. Rev. B 85, 081106 (2012), erratum 86, 199901(E) (2012)

\bibitem{19}
B. L. Altshuler, I. L. Aleiner, and V. I. Yudson, Phys. Rev. Lett. 111, 086401 (2013)

\bibitem{20}
J. I. V\"ayrynen, M. Goldstein, and L. I. Glazman, Phys. Rev. Lett. 110, 216402 (2013)

\bibitem{21}
N. Traverso Ziani, C. Fleckenstein, F. Cr\'epin, and B. Trauzettel, EPL 113, (2016)

\bibitem{22}
H-Y. Xie, H. Li, Y.-Z. Chou, and M. S. Foster, Phys. Rev. Lett. 116, 086603 (2016)

\bibitem{16}
J. C. Budich, F. Dolcini, P. Recher, and B. Trauzettel, Phys. Rev. Lett. 108, 086602 (2012)

\bibitem{12h} I. Herbut, V. Juri\v ci\'c and O. Vafek, Phys. Rev. Lett. 100, 046403 (2008)

\bibitem{13h} A. Giuliani, V. Mastropietro and M. Porta, Phys. Rev. B 83, 195401 (2011); Comm. Math. Phys. 311, 317 (2012)

\bibitem{15aaa} G. Benfatto, P. Falco, V. Mastropietro, Phys. Rev. Lett., 104, 075701, 2010; Comm. Math. Phys. 292, 569-605 (2009); Comm. Math. Phys. 330, 153--215 (2014); Comm. Math. Phys. 330, 217--282 (2014)

\bibitem{Ha} F. D. M. Haldane, Phys. Rev. Lett. 61, 2015 (1988).

\bibitem{GMP1} A. Giuliani, V. Mastropietro and M. Porta, Comm. Math. Phys. 349, 3, 1107--1161 (2016).

\bibitem{GMP2} A. Giuliani, I. Jauslin, V. Mastropietro and M. Porta, Phys. Rev. B 94, 205139 (2016).

\bibitem{Hao} N. Hao, P. Zhang, Z. Wang, W. Zhang, Y. Wang. Phys. Rev. B 78, 075438 (2008)

\bibitem{Hal} B. I. Halperin, Phys. Rev. Lett. 25, 2185 (1982).

\bibitem{Hat} Y. Hatsugai, Phys. Rev. Lett. 71, 3697 (1993).

\bibitem{SKR} H. Schulz-Baldes, J. Kellendonk and T. Richter, J. Phys. A: Math. Gen. 33, L27 (2000).

\bibitem{EG} P. Elbau and G. M. Graf, Comm. Math. Phys. 229, 415-432 (2002).

\bibitem{QWZ} X.-L. Qi, Y.-S. Wu and S.-C. Zhang, Phys. Rev. B 74, 085308 (2006).

\bibitem{AVS} J. C. Avila, H. Schulz-Baldes, C. Villegas-Blas, MPAG 16, 137--170 (2013)

\bibitem{GP} G. M. Graf and M. Porta, Comm. Math. Phys. 324, 851--895 (2013)

\bibitem{M66} G. Antinucci, V. Mastropietro, M. Porta, arXiv:1708.08517.

\bibitem{Bry} D. C. Brydges, {\it Ph\'enom\`enes critiques, syst\`emes al\'eatoires, th\'eories de jauge}, 129--183. North-Holland, %Amsterdam (1986).

%\bibitem{14h} D. L. Boyda, V. V. Braguta, M. I. Katsnelson, M. V. Ulybyshev, Phys. Rev. B 94, 085421 (2016)

\bibitem{Mas} V. Mastropietro, J. Math. Phys. 48, 022302 (2007).

\end{thebibliography}
\end{document}